\documentclass[]{article}
\usepackage[english]{babel}
\usepackage[pdftex]{graphicx}
\usepackage{enumerate}
\usepackage[tbtags]{amsmath}
\usepackage{color}
\usepackage{amssymb}
\usepackage[square, comma, numbers, sort&compress]{natbib}
\usepackage{bibentry}
\usepackage{hyperref}

\def\spsi{\vert \psi_1^{(n,j)} \rangle}
\def\spsf{\vert \psi_2^{(n,j)} \rangle}
\def\spil{\langle \psi_1^{(n,j)} \vert}
\def\spfl{\langle \psi_2^{(n,j)} \vert}
\def\sgh{\hat{\sigma}}
\def\sgm{\hat{\sigma}_-}
\def\sgp{\hat{\sigma}_+}
\def\adg{\hat{a}^{\dagger}}
\def\am{\hat{a}}
\def\Hh{\hat{H}}

\def\Evc{\vec{E}}

\def\mvc{\vec{\mu}}
\def\sgj{\vert g_j \rangle}
\def\sej{\vert e_j \rangle}
\def\snr{\vert n \rangle}
\def\snpr{\vert n+1 \rangle}
\def\sgl{\langle g_j \vert}
\def\sel{\langle e_j \vert}

\def\snpl{\langle n+1 \vert}

\begin{document}
\title{Modeling and Deciphering on Two Spin-polariton Entanglement Experiments in NV centers of diamond}
\author{Xiaodong Qi\footnote{Family name: Qi=Quantum Information, the abbreviation of the full name: QXD=Quantum X Dynamics, e-mail: qxd@physics.queensu.ca.}}\date{First Version: Jan, 2011; Available online: Nov, 2011; Third Version: Feb, 2012. }
\maketitle

\begin{abstract}
This work is a theoretical investigation on the spin-polariton (polarized single photon) entanglement in nitrogen vacancy (NV) centers in diamond in order to interpret the results of two landmark experiments~\cite{Buckley2010, Togan2010} published in {\it {Science}} and {\it {Nature}}. A Jaynes-Cummings model is applied to analyze the off- and on-resonant dynamics of the electronic spin and polarized photon system. Combined with the analysis on the NV center's electron structure and transition rules, this model consistently explained the Faraday effect, Optical Stark effect, pulse echo technology and energy level engineering technology in the way to realize the spin-polariton entanglement in diamond. All theoretical results are consistent well with the reported phenomena and data.

This essay essentially aims at applying the fundamental skills the author has learned in Quantum Optics and Nonlinear Optics, especially to the interesting materials not covered in class, in assignments and examinations, such as calculations on matrix form of Hamiltonian, quantum optical dynamics with dressed state analysis, entanglement and so on.
\end{abstract}

\section{Introduction}
\subsection{Investigation on entanglement towards photonic applications}
At the very heart of applications such as quantum cryptography, computation and teleportation lies a fascinating phenomenon known as "entanglement"--the spooky, distance-defying link that can form between objects such as atoms even when they are completely shielded from one another \cite{Dousse2010}. This type of correlation between particles is the deepest difference between Quantum and Classical world \cite{Fan2007} \cite{Eisaman2008} \cite{Cohen2009}, and is also the key to realize ``Qubits'', the unit of Quantum Information and future Quantum Computer. Because the photon is the best particle to carry information and to propagate it to a distant receiver at the fastest speed, realizing photon-photon and multi-photon entanglement is the best choice for future photonic applications \cite{Ladd2010}. Now scientists have realized two-, four-, six-, and even more photon entanglement using parametric down-conversion and other nonlinear optical technologies (see for example \cite{Eisenberg2004,Raadmark2009a}). However, it is hard to operate on photons directly, so it is necessary to study other particles' entanglement and their coupling with photons. Until now, successful quantum entanglement has also been demonstrated with individual ions and atoms (see for example this review paper \cite{Ramsay2010}), electrons in superconductors  \cite{barreiro2010experimental}, and coupling superconductor qubits to cavity light mode \cite{Majer2007}. However all of these apparatuses are either too big for application or dependent on tough conditions (like ultralow temperature) to maintain the entangled states, in other words , they are far from practical applications. Only in recent years have people made considerable progress in acceptable conditions with integrable entangled solid state systems \cite{mason2010carbon,Herrmann2010}--like semiconductor Quantum dots \cite{Muller2009}, photonic crystals \cite{Wolters2010}--and has ``spintronics'' come into the science and technology community \cite{Gregg2007} \cite{Sarchi2008}.

\begin{figure}[h]
\begin{center}
\includegraphics[width=10cm]{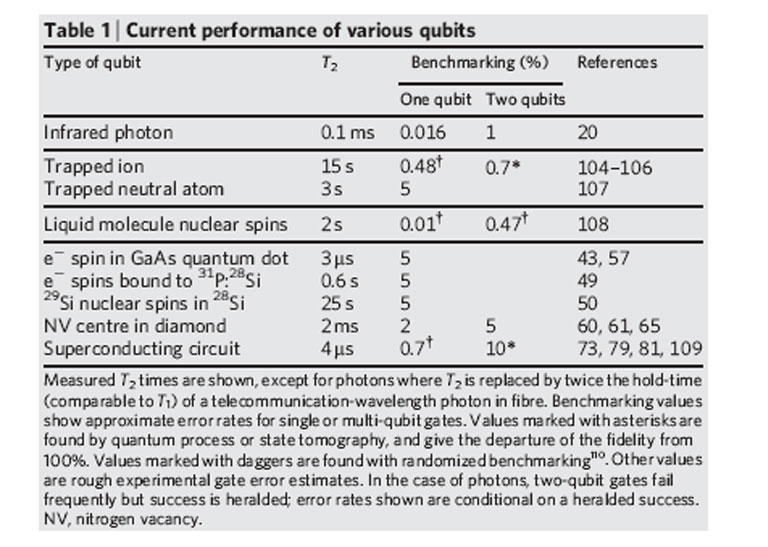}
\caption{Current performance of various qubits. Reprinted by permission from Macmillan Publishers Ltd: \href{http://www.nature.com/nature/index.html}{Nature} \textbf{464} (2010), no.~7285, 45--53\cite{Ladd2010}, copyright 2010. }\end{center}
\label{T2}
\end{figure}

It turns out electronic spin is a good carrier that can be used for computing and storing information at the same time~\cite{Sarma2001}~\cite{Sarchi2008}~\cite{Biercuk2009}~\cite{Yoo2010}. What's more, it is also easy to be coupled to polarized photons, which is believed to be a better way to enlarge our information transmitting capability without costing more energy and fibers~\cite{Ulrich1979}~\cite{Liang1999}~\cite{Olsson2000}. For the practical application in quantum memory for future computer, it both asks for long dephasing or operating time (T2) and acceptable working frequency, as well as the capability of operating in room temperature. These properties are the materials' intrinsic  properties, and will hardly be modulated along with the development of science and technology. A comparison among several promising candidates for quantum computer and quantum information applications is listed in Table.\ref{T2}.

As shown in the table NV (nitrogen vacancy) center in diamond has almost the longest operating time slot and can work from microwave to visual light band. It also can run at room temperature, owing to the Zero Phonon Line emitting \cite{Santori2010}. The probability of building a quantum computer basing on diamond and defects, like the NV center, is well confirmed by many research groups \cite{Stoneham2009,Hemmer2005,Weber2010,Ladd2010,Lambrecht2010}.  In recent years, the electron structure of NV center in diamond is widely studied using different methods \cite{Weber2010,Gali2008,Gali2009,Hossain2008,Tamarat2008,Goss1997,Lenef1997,Gonzalez2010,Larsson2008}, and the difficulties of entanglements among the spaced NV centers \cite{Bukach2010} and coupling to superconductor circuits \cite{Chen2010b}, integrated to photonic cavities \cite{Yang2010a,Yang2010b} and microrings~\cite{Faraon2011} are also overcome very recently.

However, it is a real challenge to realize the entanglement between electronic spin and polarized photon in NV centers in diamond \cite{Robledo2010} \cite{Batalov2009}. There are several difficulties. For example, the energy bands are very complex compared with individual ions and atoms exposing to the crystal environment. And it is hard to modulate the energy level structures to include two electronic spin states with almost the same energy gap to transfer to the common ground or excited states to easily operate and generate entanglement between photons and spins. In 2010, coherent operation and entanglement between electronic spins in NV center and polarized photons are performed successfully by Buckley and Togan and their teams \cite{Buckley2010} \cite{Togan2010}.

Emre Togan and collaborators reported their work in August 2010 \cite{Togan2010}, first-time showed the stable entanglement between spin state and photon polarization in a solid-state material which potentially can be demonstrated in room temperature. Buckley's work was reported in Nov 2010 \cite{Buckley2010}, on coherence control experiment for single spin in diamond, with the Faraday Effect (FE) and the Optical Stark Effect (OSE) observed under scheduled exciting pulse series. FE is usually a tool to control polarization rotating of emitted light from solid system. And OSE can cause energy band structure adjust and luminescence spectrum shift if an ultrafast (usually in femtosecond scale) laser pulse is applied on the sample. The OSE has been investigated both theoretically and experimentally in QDs of semiconductor systems (see, for example, the first experimental observation of OSE in semiconductor QD \cite{Unold2004}) and other solid optical systems \cite{Liu2009b,Sotier2009a,Vu2006}, and has been used as a tool to control single electron and photon or phonon interaction in solid nanostructures in recent years \cite{Lee2008,Lee2009}. These two effects are important for quantum operations (see for example \cite{Ramsay2010} \cite{Du2009}).

In Togan and Buckley's work, they all used the $A_2$ singlet state as the excited state and a pair of lower $m_s=\pm1$ magnetic momentum states as the ground states, and operated the system with tunable 637nm lasers. Their results show a promising way toward diamond, or more generally, solid-system based, quantum logical gates, multi-spin-photon entangled network, and other practical applications in related fields. This article will mainly focus on these two experimental reports and represent their results using a unified model.



\subsection{A brief introduction of entangled states}
Before we move onto the in-depth discussion, let's briefly introduce the basic theory of entanglement in general.

As discussed in Gerry and Knight's book \cite{Gerry2005}, the four Bell states are given by
\begin{subequations}
\begin{align}
\vert \Phi^+ \rangle &= \frac{1}{\sqrt{2}}(\vert H\rangle_A\vert H\rangle_B + \vert V\rangle_A\vert V\rangle_B), \\
\vert \Phi^- \rangle &= \frac{1}{\sqrt{2}}(\vert H\rangle_A\vert H\rangle_B - \vert V\rangle_A\vert V\rangle_B),\\
\vert \Psi^+ \rangle &=\frac{1}{\sqrt{2}}(\vert H\rangle_A\vert V\rangle_B + \vert V\rangle_A\vert H\rangle_B), \\
\vert \Psi^- \rangle &=\frac{1}{\sqrt{2}}(\vert H\rangle_A\vert V\rangle_B - \vert V\rangle_A\vert H\rangle_B),
\end{align}
\end{subequations}
where ``H'' and ``V'' are orthogonal states with ``opposite'' properties, for example, spin-up versus spin-down, polarized-to-z versus polarized-perpendicular-to-z, and so on. If a quantum system is in one of the four Bell states, then we can say this system is in a ``maximally entangled'' state. ``Maximally entangled'' means that when we trace over quantum substate B to find the density operator $\rho_A$ of quantum substate A, we obtain a multiple of the identity operator. For example, let's consider a polarized photon A and a spinning electron B are in the state of $\vert \Psi^+ \rangle$, we have
\begin{equation}
\begin{aligned}
\rho_A &=tr_B (\vert \Psi^+ \rangle \langle \Psi^+\vert)=\sum_{i=H,V}{\vert \Psi^+ \rangle \langle \Psi^+\vert i\rangle_{BB}\langle i\vert} \\
&=\frac{1}{2}(\vert H\rangle_{AA}\langle H\vert+\vert V\rangle_{AA}\langle V\vert)=\frac{1}{2}\mathbf{1}_A,
\end{aligned}
\end{equation}
similarly, $\rho_B=\frac{1}{2}\mathbf{1}_B$. This means that if we measure photon A polarized along any axis, the result is completely random, we find polarization parallel to the axis with probability
$1/2$ and polarization perpendicular to the axis with probability $1/2$. This is also true for measuring B's spin state. Therefore, if we perform any local measurement of A or B, we acquire no information about the preparation of the state, instead we merely generate a random bit (number set combined with 0 or 1).

However, when we repeat our measurements on A and B, if we get the state of A, then the state of B is acquired. Because A and B have some correlation. As Gerry and Knight's definition, the correlation function can be written as
\begin{equation}
C(H,V)=Average[A(H)B(V)],
\end{equation}
which means the average of probing A in state H while B in state V. Here, A is in H then $A(H)=1$, otherwise, A is not in H, then $A(H)=-1$. Similar to B(V). For the $\vert \Psi^+\rangle$ state's case, we can write
\begin{equation}
\label{Chv}
C(H,V) =Pr(\vert H\rangle_A \vert V\rangle_B)+Pr(\vert V\rangle_A \vert H\rangle_B)-Pr(\vert H\rangle_A \vert H\rangle_B)-Pr(\vert V\rangle_A \vert V\rangle_B),
\end{equation}
and
\begin{equation}
\begin{aligned}
Pr(\vert H\rangle_A \vert V\rangle_B)&=_A\!\!\langle H\vert _B\!\langle V\vert \, \Psi^+\rangle=1/2,\\
Pr(\vert V\rangle_A \vert H\rangle_B)&=1/2, \\
Pr(\vert H\rangle_A \vert H\rangle_B)&=0, \\
Pr(\vert V\rangle_A \vert V\rangle_B)&=0,
\end{aligned}
\end{equation}
thus that $C(H,V)=1$. That means when we know photon A is in H state, then we are sure that spin B is in state V. Because H and V is different, we call this entanglement as non-parity or antialigned entanglement. In contrast, there are parity or aligned entanglement, if A and B are always with the same type of states. This conditional probability can be measured and verified among the statistics on a large number sets of measurement experiments. If we span our measurement basis in the form of Bell states, we can get the other correlation function for conditional measurement. And if any of the correlation function gives a value greater than 0.5, then the system is in an entangled state.

For a full description of entanglement, entropy and negativity are important conceptions, you can find more details in \cite{Vidal2002}. \cite{Eisert2005} and \cite{Plenio2005} give a systematical introduction on multi-particle (bipartite, cluster, GHZ and so on) entanglement and measurement methods. And besides the discrete type (like spin states) entanglement, some continuous variables can also be entangled, like momentum, position, energy and so on. A theoretical introduction can be found in \cite{Eisert2003} \cite{Loock2000} \cite{Eisert2005a}.

Our discussion in this essay is basically on the theory and experimental methods to establish a $\vert \Psi^+ \rangle$ type entanglement between a polariton and an electronic spin in negatively charged NV center of diamond. The essay, therefore, includes the following topics: a simple but unified Quantum Optical theory to describe the polariton-spin interaction (both off-resonant and on-resonant dynamics) in NV center of diamond, the NV center's electronic structure and possible level structure to perform a good spin-polariton entanglement, represent one recent experiment--its methods and results--using our model. Through the course of theoretical analysis, I will also explain the phenomena of Faraday effect (FE) and Optical Stark effect (OSE), which are inevitable effects to consider, and point out some possible errors in two highly impacted articles.

\section{Quantum Optical theory of spin-light interaction}
From this part we mainly reference and compared with the experimental results and theoretical analysis of two up-to-date articles and one perspective article: \cite{Buckley2010}, published in Nov 2010, \cite{Togan2010}, published in Aug 2010, and \cite{Milburn2010} as a comment on \cite{Buckley2010}, published in {\textit {Science}} in Nov 2010. I will build up a unified theory frame mainly inspired by Buckley's article, using the knowledge I have learned through the course, to explain the reported phenomena of Faraday effect (FE), Optical Stark effect (OSE), spin echo and entanglement generation between nitrogen vacancy center electronic spin and photon, and to point out some new discoveries and argument through comparing my calculation with published results. To make our discussion well readable, some contents are cited from the original papers without notice, and some parts will be highlighted with colored fonts to distinguish my disparate arguments with the publication.

\subsection{Off-resonant dynamics theory and FE, OSE in diamond}
\begin{figure}[h]
\begin{center}
\includegraphics[width=8cm]{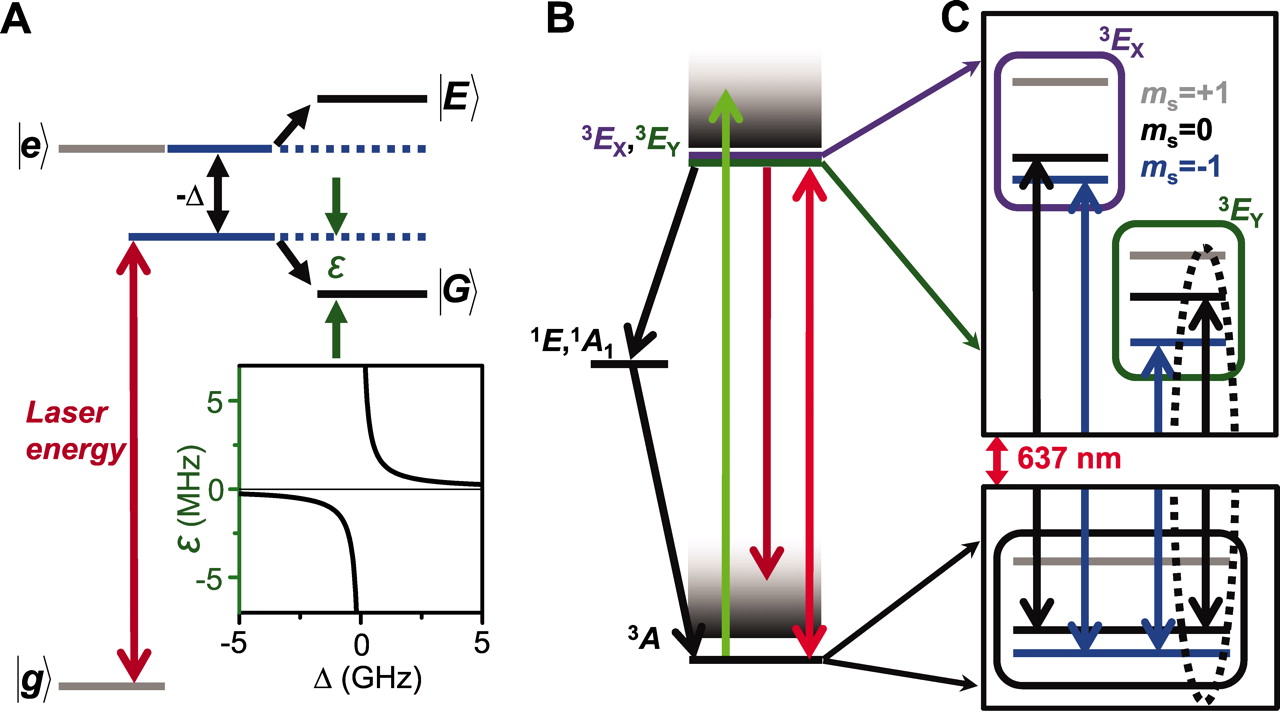}
\end{center}
\caption{Level structure for spin-photon interaction in diamond using Jaynes-Cummings model \cite{Buckley2010}. \textbf{A} shows the simplified levels and energy shift with light-electron interaction. \textbf{B,C} show the detailed level structure in reality. From \href{http://www.sciencemag.org/content/330/6008/1212.short}{B.~Buckley, G.~Fuchs, L.~Bassett and D.~Awschalom. Spin-Light Coherence for Single-Spin Measurement and Control in Diamond \textit{Science}, \textbf{330} (2010), 1212}\cite{Buckley2010}. Reprinted with permission from AAAS.}
\label{JClevel}
\end{figure}

Now let's consider the nitrogen-vacancy center under a coherent laser pump and excited from ground state to excited state.
The initial and final states for the joint spin-photon system can be written as
\begin{subequations}
\begin{align}
\label{phi0}
\vert \psi_1^{(n,j)} \rangle &= \vert g_j \rangle \vert n+1 \rangle \\
\vert \psi_2^{(n,j)} \rangle &= \vert e_j \rangle \vert n \rangle,
\end{align}
\end{subequations}
where $\sgj (\sej)$ are the bare ground (excited) states of the NV center orbital transition to spin number $m_s=j, \, j=-1,0,+1$ state, and $\snr$ is a photon-number state of the electromagnetic field, $n=0,1,2,\cdots$. As the energy splitting between different $m_s$ states is so small compared with the detuning, we can safely simplify the interested energy levels into a two-level system, with a formalism similar to Jaynes-Cummings model. The energy level structure is shown in Fig.\ref{JClevel}.

Now we can try using dipole approximation to write the interaction Hamiltonian as
\begin{subequations}
\label={Hint}
\begin{align}
\hat{H}_{int} &= C\hat{\vec{\mu}} \hat{\vec{E}} \\
 &= i\frac{\hbar\Omega_0}{2}(\hat{a}^{\dagger}  \hat{\sigma}_--\hat{a} \hat{\sigma}_+),
 \end{align}
\end{subequations}
where I used $\hat{\vec{\mu}}= \left|\vec{\mu} \right| \left(\hat{\sigma}_++\hat{\sigma}_- \right)$ and $\hat{\vec{E}}=i \left| \vec{\mu} \right| \left( \hat{a}^{\dagger}+\hat{a} \right)$ for the electric field and optical dipole, respectively, and $\hat{a}^{\dagger} (\hat{a})$ and $\hat{\sigma}_+ (\hat{\sigma}_-)$ are creation (annihilation) operators for optical photons and NV center excitations, respectively, and we neglect energy-nonconserving terms
${\hat{a} \hat{\sigma}_-},\hat{a}^{\dagger}\hat{\sigma}_+$ as a result of rotating wave approximation. For this case, I also make $C=F_{DW}^2$, $F_{DW}=0.04\pm0.01$ is the Debye-Waller factor which empirically accounts for the reduced resonant coupling between
NV center ground and excited states due to non-resonant phonon-assisted transitions. And
\begin{equation}
\label{moment2}
\left| \vec{\mu} \right|^2=\frac{3\pi\varepsilon_0\hbar^4 c^3 \gamma}{E_{ph}^3n_D},
\end{equation}
where $E_{ph}=1.945 \, eV$ is the photon energy and $n_D=2.4$ is the refractive index of diamond. We also used the on-resonance optical Rabi frequency $\Omega_0$ as
\begin{equation}
\label{Omega0}
\Omega_0=\frac{\sqrt{F_{DW}} }{\hbar}|\mvc | | \Evc_0 | \cos(\delta),
\end{equation}
where $\cos(\delta)$ accounts for the geometric coupling between the NV center dipole and the linearly polarized light.

Considering the non-interacting Hamiltonian for the spin and light field, given by
\begin{equation}
\label{H0}
\hat{H}_0=E_{ph}(\adg \am + \frac{1}{2})+E_j \frac{\sgh_z}{2},
\end{equation}
where $E_j$ is the transition energy for the spin state with spin number $m_s=j$, $\sgh_z=\sej \sel-\sgj \sgl$ and we have ignored the zero-field energy and make the average spin energy to be zero for simplicity, we can finally write the Jaynes-Cummings-like Hamiltonian for the system as
\begin{subequations}
\begin{align}
\label{Hjcdef}
\hat{H}_{JC} &=\Hh_0+\Hh_{int} \\
&= E_{ph}\adg\am+E_j\frac{\sgh_z}{2}+\frac{\hbar\Omega_0}{2} (\am\sgp + \adg\sgm).
\end{align}
\end{subequations}
Now let us rewrite the Hamiltonian into matrix formalism. Using the following relationships:
\begin{subequations}
\begin{align}
\sgp &=\sej\sgl, \label{sigma+} \\
\sgm &=\sgj\sel, \label{sigma-} \\
\sgh_z &=\sej\sel-\sgj\sgl, \label{sigmaz} \\
\am\snr &=\sqrt{n}\vert n-1 \rangle, \label{an} \\
\adg\snr &=\sqrt{n+1}\vert n+1 \rangle, \label{adgn}
\end{align}
\end{subequations}
we get the elements for the Hamiltonian matrix as
\begin{subequations}
\begin{align}
H_{11}^{(n,j)} &= \langle \psi_1^{(n,j)} \vert \Hh_{JC} \spsi, \nonumber \\
&= \snpl \sgl \, [E_{ph}\adg\am+E_j\frac{\sgh_z}{2}+ \frac{\hbar\Omega_0}{2}(\am \sgp + \adg \sgm)] \, \sgj \snpr, \nonumber \\
&= E_{ph}(n+1)-\frac{E_j}{2},
\label{H11} \\
H_{12}^{(n,j)} &=\spil \Hh_{JC} \spsf =\frac{\hbar \Omega_0}{2} \sqrt{n+1},
\label{H12} \\
H_{21}^{(n,j)} &=\spfl \Hh_{JC} \spsi =\frac{\hbar \Omega_0}{2} \sqrt{n+1},
\label{H21} \\
H_{22}^{(n,j)} &=\spfl \Hh_{JC} \spsf =E_{ph}n+\frac{E_j}{2}.
\label{H22}
\end{align}
\end{subequations}
Hence the Hamiltonian matrix gives
\begin{equation}
\begin{aligned}
\hat{\mathbf{H}}^{(n,j)}_{JC} &= \left( \!
\begin{array}{ccc}
E_{ph}(n+1)-\frac{E_j}{2} &\quad \frac{\hbar \Omega_0}{2} \sqrt{n+1}\\
 \frac{\hbar \Omega_0}{2} \sqrt{n+1} &\quad E_{ph}n+\frac{E_j}{2}
\end{array} \! \right), \\
&= \left( \! \!
\begin{array}{ccc}
E_{ph}(n+ \frac{1}{2})+ \frac{\hbar\Delta_j }{2} &\quad
\frac{\hbar \Omega_0}{2} \sqrt{n+1} \\
\frac{\hbar \Omega_0}{2} \sqrt{n+1} &\quad
E_{ph}(n+ \frac{1}{2})- \frac{\hbar\Delta_j}{2}
\end{array} \! \right) \label{Hmatrix},
\end{aligned}
\end{equation}
where I made $\Delta_j=\left( E_{ph}-E_j \right)/\hbar$ as the detuning of the laser from the unshifted NV center transition frequency.
Now the eigenequation gives
\begin{equation}
\hat{\mathbf{H}}_{JC}^{(n,j)}\vert \Psi^{(n,j)} \rangle=E(n,\Delta_j)\vert \Psi^{(n,j)} \rangle,
\end{equation}
where $E(n,\Delta_j)$ is the eigenvalue or eigenenergies of the system and $\vert \Psi^{(n,j)} \rangle$ stands for the eigenstate or eigenvector in basis of $\spsi$ and $\spsf$. The equation has a nontrivial solution only if
\begin{equation}
\mathbf{Det}\left(\hat{\mathbf{H}}_{JC}^{(n,j)}\right)=0,
\end{equation}
which gives the two eigenenergies as
\begin{equation}
\begin{split}
E_\pm(n,\Delta_j)=& E_{ph}(n+ \frac{1}{2}) \pm \frac{\hbar}{2} \sqrt{\Delta_j^2 + \Omega_0^2(n+1)},\\
=& E_{ph}(n+ \frac{1}{2}) \pm \frac{\hbar}{2} \sqrt{\Delta_j^2 + \Omega_n^2}.
\label{eigenE}
\end{split}
\end{equation}
Notice that in Equ.~\ref{eigenE}, I have defined the $n$-photon on-resonance optical Rabi frequency for a pulse with fixed duration as
\begin{equation}
\label{Omega_n}
\Omega_n=\Omega_0\sqrt{n+1},
\end{equation}
so that $\Omega_n$, here, has the same physics meaning as $\Omega_0$ in the Equs.~S11 and~S9 of paper \cite{Buckley2010}, which can be observed experimentally (thank the authors of Ref.~\cite{Buckley2010} for the correction on my first version of this article). Now, $\Omega_0$ gives the $per$-photon atom-photon coupling Rabi frequency. As you will see later, the experimental analysis in Ref.~\cite{Buckley2010} is consistent with this coupling model.

If we substitute the eigenenergies into the eigenequation respectively, we can solve the equation and normalize the results to obtain the eigenvectors finally, which depict the states of the system. We write the eigenstates, associated with eigenenergies $E_{\pm}(n,\Delta_j)$, as dressed states, which read
\begin{subequations}
\begin{align}
\vert n,+j\rangle &= \cos(\Phi_{nj}/2)\vert \psi_1^{(n,j)}\rangle +\sin(\Phi_{nj}/2)\spsf, \label{dressstates1}\\
\vert n,-j\rangle &= -\sin(\Phi_{nj}/2)\spsi + \cos(\Phi_{nj}/2)\spsf, \label{dressstates2}
\end{align}
\end{subequations}
where $j$ is just a label to spin states and the phase factor $\Phi_{nj}$ is defined as
\begin{equation}
\Phi_{nj} = \arctan \left( \frac{\Omega_n}{\Delta_j}\right),
\end{equation}
with
\begin{subequations}
\begin{align}
\sin(\Phi_{nj}/2) &= \frac{1}{\sqrt{2}}\left[ 1-\frac{\Delta_j}{\sqrt{\Delta_j^2+\Omega_n^2}} \right]^{1/2}, \\
\cos(\Phi_{nj}/2) &= \frac{1}{\sqrt{2}}\left[ 1+\frac{\Delta_j}{\sqrt{\Delta_j^2+\Omega_n^2}} \right]^{1/2}.
\end{align}
\end{subequations}
To get a general dynamic solution for the spin-photon system, we suppose the field is initially prepared in a superposition of number states
\begin{equation}
\vert \Psi_f(0)\rangle =\sum_{n,j}{C_{n,j}\snr},
\end{equation}
and the electron is in excited state $\vert e_j\rangle$. Thus the initial state for the whole system is
\begin{equation}
\vert \Psi(0)\rangle =\sum_j{\vert \Psi_f(0)\rangle \sej} = \sum_{n,j}{C_{n,j}\snr \sej}=\sum_{n,j}{C_n\spsi}.
\end{equation}
Since Eqs. \ref{dressstates1} and \ref{dressstates2} give
\begin{equation}
\spsi = \cos(\Phi_{nj}/2)\vert n, +j \rangle-\sin(\Phi_{nj}/2)\vert n,-j\rangle,
\end{equation}
thus the initial states for the system can be described through dressed states as
\begin{equation}
\vert \Psi(0)\rangle = \sum_{n,j}{C_{n,j}[\cos(\Phi_{nj}/2)\vert n,+j\rangle -\sin(\Phi_{nj}/2)\vert n,-j\rangle]}.
\end{equation}
Because the derivation for dressed states above is in the Heisenberg picture, the states are time independent. If we transfer them into the Schrodinger picture, we can easily get the dynamic state vector for times $t>0$ as
\begin{equation}
\begin{aligned}
&\quad \vert \Psi (t) \rangle = \exp\left[-\frac{i}{\hbar}\hat{H} t\right] \vert \Psi(0)\rangle, \\
&= \sum_{n,j}{C_{n,j} \left[ \cos(\Phi_{nj}/2) e^{-iE_+ (n,\Delta_j) t/\hbar} \vert n,+j\rangle  - \sin(\Phi_{nj}/2) e^{-iE_-(n,\Delta_j) t/\hbar} \vert n,-j\rangle \right]}.
\end{aligned}
\label{dynamics}
\end{equation}

Now let us move on to analyze the energy level shift as a characteristic of OSE.

Equation (\ref{eigenE}) shows the energy splitting between the two eigenenergies are
\begin{equation}
\label{Esplit}
E_{split}(n,\Delta_j)= E_+(n,\Delta_j)-E_-(n,\Delta_j)=
\hbar \sqrt{\Delta_j^2+\Omega^2_n}.
\end{equation}
If $\Delta_j=0$ or no detuning, the splitting energy is $E_{split}(n,0)=\hbar \Omega_0\sqrt{n+1}=\hbar \Omega_n$ (we call it the Rabi energy for system with $n$ photons, $\Omega_n$ is the Rabi angular frequency for this system), corresponding to the spin splitting between different $m_s$ states (here charaterized by $\Omega_0$) and associated with photon number state $n$. If $\Delta_j \neq 0$ the energy splitting will increase, and $E_{\pm}$ moves up or down and becomes the polariton eigenenergy for $\Delta_j >0$ or $\Delta_j < 0$ to make the maximum overlap with the initial state $\vert \psi_0 \rangle$.

Suppose initially the system is in ground state with total energy
\begin{equation}
\label{Eg0}
E_{g0}=E_{ph}(n+1)-\frac{E_j}{2}=E_{ph}(n+\frac{1}{2}) + \frac{\hbar \Delta_j}{2}.
\end{equation}
After the detuning laser interacting with the NV center free-electron, the system redistributes its eigenenergies (observed energies) to $E_{\pm}(n,\Delta_j)$. Now let's consider the case that $|\Delta_j| \gg \Omega_n$ and the photon's or incident light's energy is below the upper level or $\Delta_j<0$, which means the excited states will occupy the $E_-(n,\Delta_j)$ level.
From equation (\ref{eigenE}), we can get the energy shift as
\begin{equation}
\label{Eg}
\begin{aligned}
\varepsilon_g(n,\Delta_j) &=E_-(n,\Delta_j) - E_{g0}=\frac{\hbar}{2}(\Delta_j-\sqrt{\Delta_j^2+\Omega_n^2}) \\
&= \frac{\hbar| \Delta_j|}{2} \left[\sqrt{1+\frac{\Omega_n^2}{\Delta_j^2}}-1 \right] \\
&\approx \frac{\hbar}{4}\frac{\Omega_n^2}{|\Delta_j|},
\end{aligned}
\end{equation}
corresponding to Equ.~S13 in Ref.~\cite{Buckley2010}.

Now let's consider the photon number and accumulated effects for a pumping pulse with fixed duration. The pulse duration $\tau $ in paper \cite{Buckley2010} $\approx 1\mu s$ with power $\approx 1 \mu W$ only contains $\approx 10^6$ photons with wavelength $\approx 637 nm$ and detuning in GHz range. The energy shift during the pulse lightening yields a net phase shift to the polariton, given by
\begin{equation}
\phi(n,\Delta_j)=\frac{\tau \varepsilon_g}{\hbar} \approx \frac{\tau \Omega_n^2 }{4\Delta_j} \approx D\frac{n}{\Delta_j},
\label{phaseshift}
\end{equation}
where
\begin{equation}
\label{D}
D=\frac{|\vec{\mu}|^2F_{DW}E_{ph}\cos^2(\delta)} {2\hbar^2cn_D\varepsilon_0A_{eff}},
\end{equation}
considering vacuum electric field
\begin{equation}
\label{E0}
|\vec{E}_0|= \sqrt{\frac{2nE_{ph}}{n_D\varepsilon_0A_{eff}c\tau}},
\end{equation}
since
\begin{equation}
\label{I}
I=\frac{cn_D\varepsilon_0}{2}|\vec{E}_0|^2
=\frac{nE_{ph}}{\tau A_{eff}}.
\end{equation}
We can estimate the accumulated phase shift for one photon is $\phi_j=D/\Delta_j \approx 10^{-5}\, rad$, or  which is consistent with the experimental result in Ref.~\cite{Buckley2010} for $D/2\pi \approx 10 kHz$. So for the bunch of photons in one pulse, we can obtain an observable signal phase from the accumulated phase in the order of $\mu rad$, as correctly analyzed in Ref.~\cite{Buckley2010} by the authors. This phase change corresponds to the rotation of the polarized output light, so it shows the Faraday effect.

To gain the exact expression for both the FE and the OSE, the authors used coherent state $\vert \alpha \rangle$ to describe the laser field--which is valid for such laser input with considerable photon in one pulse--and calculated the reduced density matrices, $\hat{\rho}_{spin}$ and $\hat{\rho}_{light}$, for spin and optical components. Now suppose the initial state of the polariton is
\begin{equation}
\label{Psi}
\vert \Psi_0 \rangle = ( \sum_j{\beta_j \vert g_j \rangle} ) \vert \alpha \rangle.
\end{equation}
As introduced in class, the coherent state gives
\begin{equation}
\label{alpha}
\vert\alpha\rangle=e^{-\frac{|\alpha|^2}{2}}\sum_n{\frac{\alpha^n}{\sqrt{n!}}
\vert n \rangle},
\end{equation}
 with $\alpha=|\alpha| e^{i\gamma}$ and $|\alpha|^2=\langle n \rangle$ bearing the mean number of photons, such that the polariton evolves to the state
\begin{subequations}
 \begin{align}
\vert\Psi\rangle
&= \sum_j{\beta_j e^{-\frac{|\alpha|^2}{2}} \sum_n{\frac{\alpha^n}{\sqrt{n!}} e^{i \Phi (n,\Delta_j)} \sgj \vert n \rangle}}, \\
 &= \sum_j{\beta_je^{-\frac{|\alpha|^2}{2}}
\sum_n{\frac{(\alpha e^{i \phi_j})^n}{\sqrt{n!}}
\sgj \vert n \rangle}}, \\
 &= \sum_j{\beta_j \sgj \vert \alpha e^{i \phi_j} \rangle}, \label{Psif}
 \end{align}
\end{subequations}
where $\phi_j$ is the accumulated phase per photon by the state $\sgj \vert \alpha \rangle$. From the full density matrix of the resulting spin-light system, given by $\hat{\rho}=\vert \Psi \rangle \langle \Psi \vert$, the reduced density matrix for the optical field gives
\begin{subequations}
\label{rholight}
\begin{align}
\hat{\rho}_{light} &= \sum_j{\sgl \hat{\rho} \sgj}
= \sum_j{|\beta_j|^2 \vert \alpha e^{i \phi_j} \rangle} \langle \alpha e^{i \phi_j}\vert, \\
&= \vert light \rangle \langle light \vert,
\end{align}
\end{subequations}
with the ``light'' states
\begin{equation}
\label{Slight}
\vert light \rangle = \sum_j{\beta_j \vert \alpha e^{i \phi_j} \rangle},
\end{equation}
where $\beta_j=|\beta_j|e^{i\phi_{\beta_j}}$ and $|\beta_j|^2$ gives the possibility of the field occupies the spin state $\sgj$ (we will discuss this for detail later).

Similarly, the reduced spin density matrix gives
\begin{subequations}
\label{rhospin}
\begin{align}
\hat{\rho}_{spin} &= \langle \alpha  \vert \hat{\rho} \vert \alpha \rangle = \sum_{j,k}{\beta_k^*\beta_j \exp{\{-|\alpha|^2[(1-e^{i \phi_j})+(1-e^{-i \phi_k})]\}} \, \sgj \vert g_k \rangle} \\
&\approx \sum_{j,k}{\beta_k^*\beta_j \exp{[-i\langle n \rangle( \phi_j- \phi_k)]} \, \sgj \vert g_k \rangle}, \\
&= \vert spin \rangle \langle spin \vert.
\end{align}
\end{subequations}
In the last approximation, I used Taylor expansion of $e^{i\phi_j}$ and $e^{-i\phi_k}$, since $\phi_j \ll 1$, and hence the ``spin'' states become
\begin{equation}
\label{Sspin}
\vert spin \rangle =\sum_j{\beta_j e^{i\langle n\rangle\phi_j} \sgj}= \sum_j{\beta_j e^{i \frac{\tau \Omega_n^2}{4\Delta_j}}}.
\end{equation}
From the above equations, we can see that the rotated spin states are affected mainly by the detuning ($\Delta_j$) from the light field for an given pulse and occupation possibilities on $j$ states (associated with $\beta_j$, which is determined by the initially excited states using on-resonance echo technology which will be introduced in later sections.
Since the spin probability amplitudes $\beta_j$ and phase information ($\phi_j$) are passed on to the optical field, which is described by equation (\ref{Slight}), through measuring the optical field's information, we can get the complete information on the spin-photon interaction system. This technology is named as ``spin-light coherence for single-spin measurement and control'' technology in Ref.~\cite{Buckley2010}.

Now let's consider the optical property, to build up the bridge linking between quantum optical theory and experimental measurement, basing on ``light'' states or reduced density matrix for the optical field. Assuming the electric field operator gives
\begin{equation}
\hat{\vec{E}}(\vec{r},t)= |\vec{E}_0|\vec{u}(\vec{r}) e^{- i \omega t}(\hat{a}+\adg),
\end{equation}
where $\vec{u}(\vec{r})$ describes the spatial mode of the optical field. The expectation value of the field gives
\begin{equation}
\begin{aligned}
\langle \hat{\vec{E}}(\vec{r},t)\rangle_{light} &= \langle light \vert \hat{\vec{E}}(\vec{r},t) \vert light \rangle \\
&= |\vec{E}_0|\vec{u}(\vec{r}) e^{- i\omega t}|\alpha| \sum_j{[\beta_j e^{i(\phi_j+\gamma)} + \beta_j^* e^{-i(\phi_j+\gamma)}]} \\
&= 2|\vec{E}_0|\vec{u}(\vec{r}) e^{- i\omega t}|\alpha| \sum_j{|\beta_j|cos(\phi_j'+\gamma)},
\end{aligned}
\end{equation}
where  $\phi_j'=\phi_j+\phi_{\beta_j}$ is the phase corresponding to the $m_s=j$ spin state. Since phase information associated with $\omega$ is independent to the amplitude of the expected field value, and also $\omega$ is a very large number ($\approx 10^{15} \, rad/s$), $\phi_j'$ dominates the observable phase information. Meanwhile, because $\phi_{\beta_j}$ only depends on the initial state, which can be conditioned into some fixed state, $\phi_j'$ is strongly dependent on $\phi_j=D/\Delta_j$ which is basically a function of detuning $\Delta_j$. And this phase information varies the magnitude of the measured field strength or output light density considerably
(unfortunately, the authors did not provide this measurement data in their paper). In the far off resonant detuning experiment, only one linear polarization component of light is coupled to the transition channel $j$, hence the polarized phase is shifted relative to the non-interacting polarization state by an amount $\phi_j$, which can be measured by using an adjustable polarizing beam splitter. And the Faraday phase $\Phi_F$ is the difference in phase between the $m_s=0$ and $m_s=-1$ (here states $m_s=\pm 1$ are almost degenerate compared with $m_s=0$ state) spin state. As the photon distribution on different polariton states is related with the electron distribution on different spin states (described by $\beta_j$),
so that we can bring in Faraday phase $\Phi_F(N)$ for a system with total number of $N$, by
\begin{equation}
\Phi_F(N)=\phi_0'-\phi_{-1}=D(\frac{1}{\Delta_0}-\frac{1}{\Delta_{-1}})= -D\frac{\omega_s }{\Delta_0 \Delta_{-1}},
\end{equation}
where $\omega_s=(E_{-1}-E_0)/\hbar$ is the frequency spacing between the resonances to $m_s=0,-1$ states.
Assuming the high-order atom-photon interaction is negligible, the effect Faraday phase is almost a constant representing the $per$-photon effect for a system with a total photon number of $N$.

\begin{figure}[h]
\begin{center}
\includegraphics[width=7.5cm]{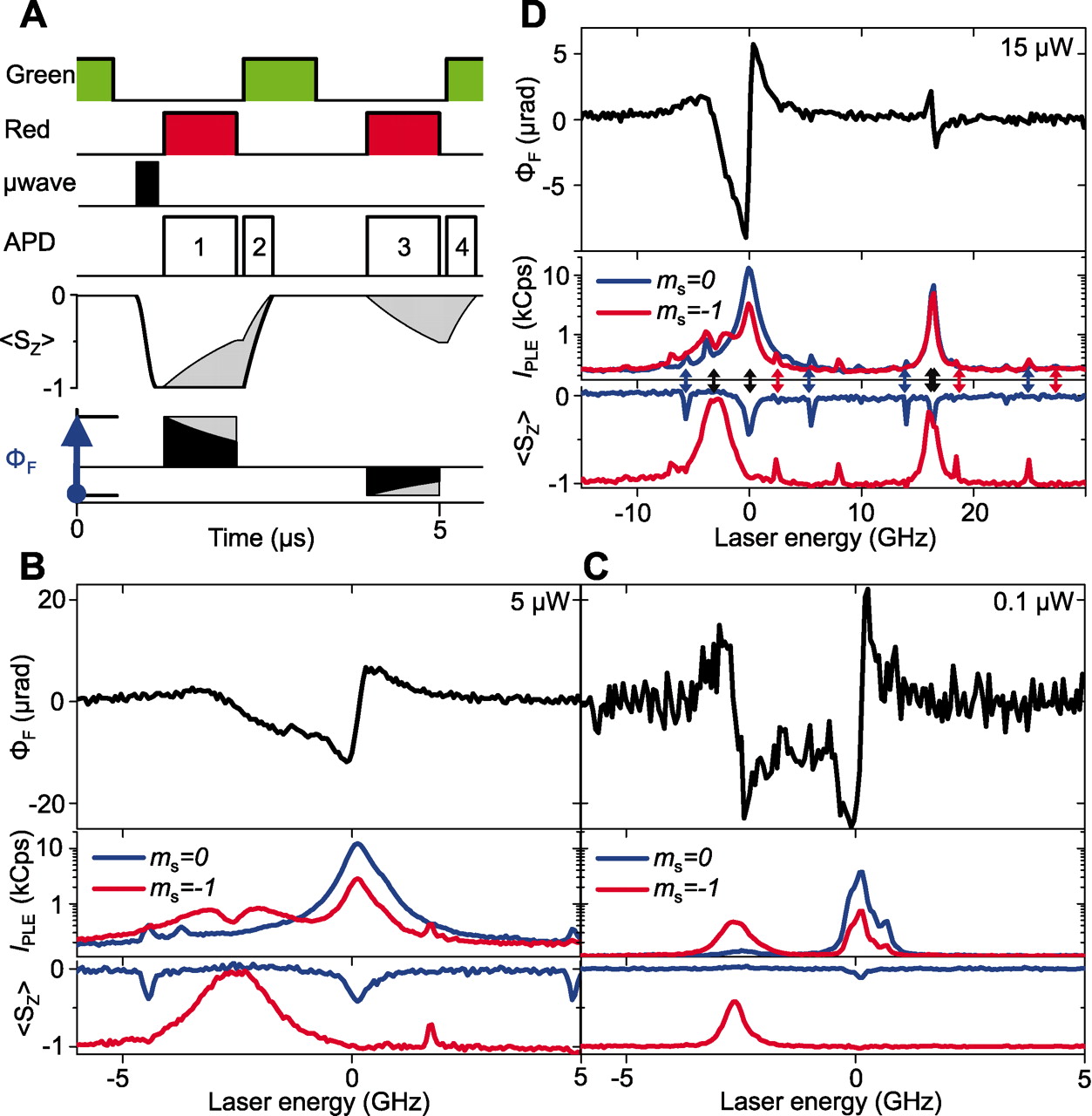}
\end{center}
\caption{\textbf{(A)}FE measurement timing sequence show the gated green laser, tunable red laser,
microwave, and multiplexed APD timing, as well as the spin-state evolution and diode bridge signal in
time. PL photons are binned separately in time to measure both $I_{PLE}$ (APD measured, with Lorentz fit) and $\langle S_Z\rangle$ (the Larmor precession rate shift due to the polarization energy shift, given by $S_j=\varepsilon_g(\langle n\rangle/2\pi\hbar,)$) for the $m_s =-1$ and $m_s = 0$ prepared spin states. Gray areas indicate possible spin-polarization effects
from the red laser. $\Phi_F$ (denoted by the blue arrow at bottom-left) is the difference in measured red laser
polarization response between the prepared $m_s = 0$ and $m_s=-1$ spin states. \textbf{(B and C)} FE data sets scaled
identically show $\Phi_F$, as well as $I_{PLE}$ and $\langle S_Z \rangle$, for both $m_s =0$ (blue) and $m_s =-1$ (red) prepared spin states of the $^3E_Y$ orbital-branch optical transitions at 5 and $0.1$mW red laser power, respectively. The $m_s = 0$ optical transition (at $0$GHz laser energy) is more robust against spin polarization than the $m_s =-1$ optical
transition (near -3$GHz$ laser energy). $\mu$rad, microradians; $kCps$, photon kilocounts per second. \textbf{(D)} FE
data set at $15$mW scans across both $^3E_X$ and $^3E_Y$ orbital-branch optical transition energies. The measured FE is substantially reduced for the $^3E_X$ orbital transitions (near $16.5$GHz), primarily as a result of
the smaller optical transition energy splitting between the spin states. From \href{http://www.sciencemag.org/content/330/6008/1212.short}{B.~Buckley, G.~Fuchs, L.~Bassett and D.~Awschalom. Spin-Light Coherence for Single-Spin Measurement and Control in Diamond \textit{Science}, \textbf{330} (2010), 1212}\cite{Buckley2010}. Reprinted with permission from AAAS. }
\label{FE}
\end{figure}

 Since $I \propto N$, one expects to observe the FE effective phase shifting with various input light power. This experiment shows a good agreement with the expectation as in figure (\ref{FE}).

By accumulating Faraday phase in the total coherent optical field with N photons, we can get the relative OSE phase shift as
\begin{equation}
\label{phaseOSE}
\Phi_{OSE}\approx N \Phi_F.
\end{equation}
And the corresponding OSE frequency shift is
\begin{equation}
\label{freqOSE}
\Sigma_S=\frac{\Phi_{OSE}}{2\pi \tau}=\frac{P_L}{4 \pi E_{ph}}\Phi_F,
\end{equation}
where $P_L=2nE_{ph}/\tau$. So the OSE phase or frequency shift is proportional to the total laser power.

\begin{figure}[h]
\begin{center}
\includegraphics[width=7.5cm]{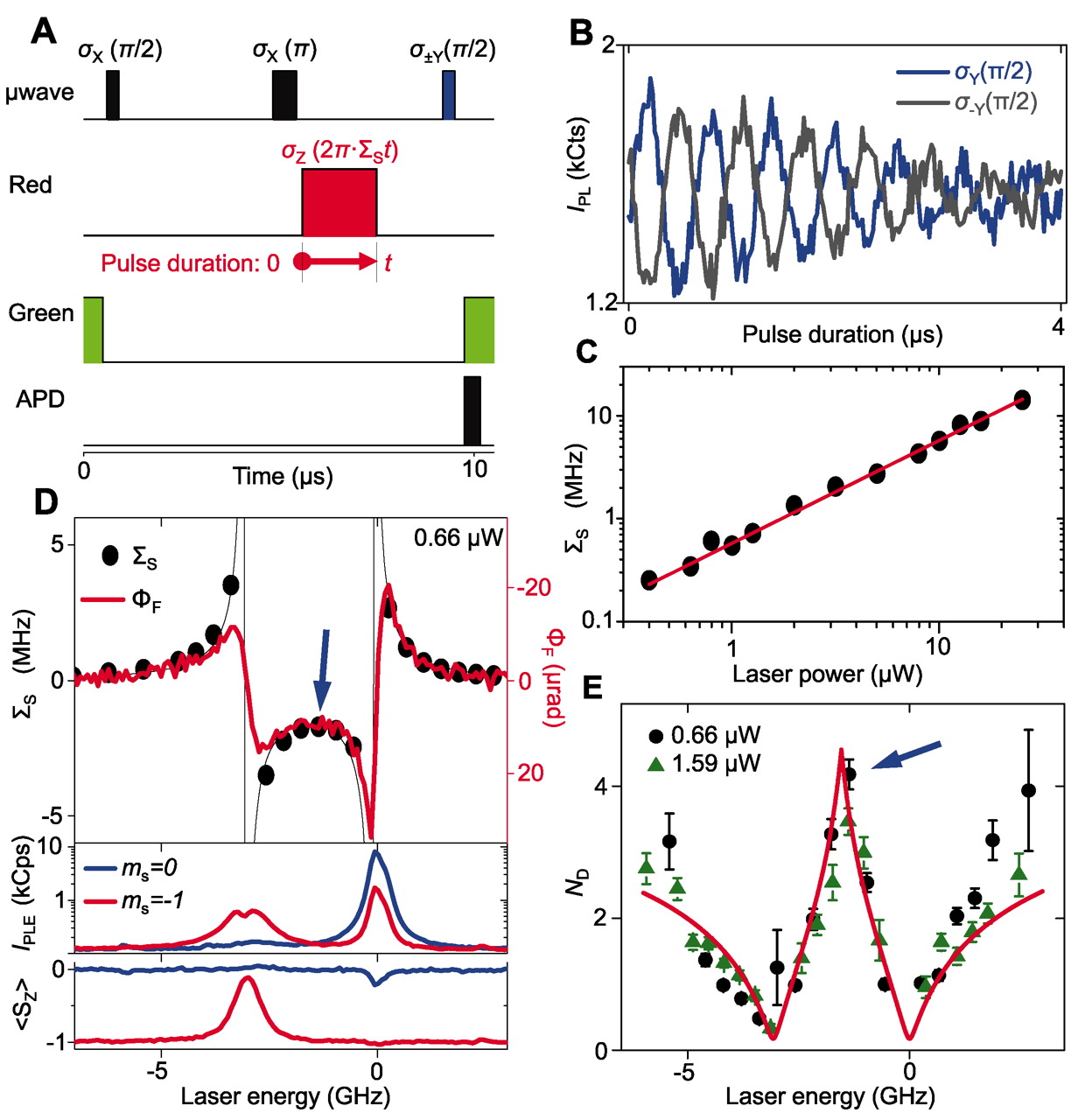}
\end{center}
\caption{ \textbf{(A)} OSE measurement timing sequence. The red laser produces a $\sigma_Z$ spin rotation proportional to
its $0$ to $4\mu$s pulse duration. $\langle S_Z\rangle$ follows $\pm \sin(\sigma_Z)$ as a result of a $\pm 90^{\circ}$ microwave phase shift of the final Hahn echo pulse. \textbf{(B)} $I_{PL}$ of sequences that have these two microwave phases. $kCts$, photon kilocounts. \textbf{(C)}
$\Sigma_S$ as a function of laser power at $\sim$ $2$GHz laser energy. The error is represented by the data-point size and
is dominated by the 15\% uncertainty in the laser power calibration. The red line is a linear fit of $\Sigma_S$ with $5.74$MHz per microwatt slope. \textbf{(D)} Comparison of $\Sigma_S$ and $\Phi_F$ showing their complimentary response as a
function of laser energy and also showing $I_{PLE}$ and $\langle S_Z\rangle$ taken with $\Phi_F$ for both prepared spin states. OSE
and FE data sets were taken under the same experimental conditions at 0.66-$\mu W$ laser power. The $\Sigma_S$ data point denoted by the blue arrow is the frequency fit of the data presented in Fig.3B. \textbf{(E)} OSE spin coherence measured in number ($\textbf{N}_D$) of $\sigma _Z$ rotations to the $1/e$ decay point. Black circles represent the measurements with $0.66\mu$W red laser power (Fig.~D), and green triangles represent measurements with
$1.59\mu$W red laser power. The blue arrow denotes the $\mathbf{N}_D$ data point fit from in Fig.~B. The red line is a fit to the data using a model incorporating dephasing from spectral broadening and laser intensity fluctuations. Error bars indicate standard errors of OSE data fits. From \href{http://www.sciencemag.org/content/330/6008/1212.short}{B.~Buckley, G.~Fuchs, L.~Bassett and D.~Awschalom. Spin-Light Coherence for Single-Spin Measurement and Control in Diamond \textit{Science}, \textbf{330}(2010), 1212}\cite{Buckley2010}. Reprinted with permission from AAAS.}
\label{OSE}
\end{figure}

By making the full FE line an odd Lorentzian curve, as a result of the Kramers-Kronig relation, and considering the dephasing mechanism is dominated by spectral diffusion and power fluctuation, the authors also analyzed the dephasing phenomena in experiments, which is quantized by $\textbf{N}_D$ that is defined as the number of OSE-induced oscillations at which
the amplitude envelope drops to 1/e times its value at $\tau=0$. The experimental data agrees with the theoretical model quite well as shown in Fig.\ref{OSE}.

\subsection{On-resonant dynamics theory and pulse echo technology}
Now let's specify our case in $\Delta_j=0$, that is when the light is resonant with the spinning electrons. Now $\Phi_n=\pi/2$, and
\begin{equation}
\begin{aligned}
E_\pm(n,j) &= E_\pm(n,\Delta_j=0)=E_{ph}(n+\frac{1}{2})\pm\frac{\hbar\Omega_0}{2} \sqrt{n+1} \\
&= E_{ph}(n+\frac{1}{2})\pm\frac{\hbar\Omega_n}{2},
\end{aligned}
\end{equation}
Eq. (\ref{dynamics}) gives
\begin{equation}
\begin{aligned}
\vert \Psi (t)\rangle &= \frac{1}{\sqrt{2}}\sum_{n,j}{C_{n,j}\left[ e^{-iE_+(n)t/\hbar }\vert n,+j\rangle-e^{-iE_-(n)t/\hbar }\vert n,-j\rangle\right]} \\
&= \frac{1}{2}\sum_{n,j}{C_{n,j}\left[ e^{-iE_+(n)t/\hbar }(\spsi+\spsf) -e^{-iE_-(n)t/\hbar }(\spsf-\spsi)\right]} \\
&=\sum_{n,j}{e^{-iE_{ph}(n+\frac{1}{2})t/\hbar}C_{n,j}\left[\cos(\frac{\Omega_0t}{2})\spsi
-i\sin(\frac{\Omega_0t}{2})\spsf \right]}.
\end{aligned}
\end{equation}
According to Eq. (\ref{phi0}), for the $n$ photon number state, at time $t$ the probability of finding the system in $\vert \Phi_1^{(n,j)} \rangle$ state is
\begin{equation}
\begin{aligned}
P_1(n,j,t) &=\left| \langle \Phi_1^{(n,j)}\vert \Psi(t)\rangle\right|^2 \\
&=\left| \sum_{m,k}{e^{-iE_{ph}(m+\frac{1}{2})t/\hbar}C_{m,k}\cos(\frac{\Omega_0t}{2})\langle \Phi_1^{(n,j)} \vert \Phi_1^{(m,k)}\rangle \delta_{m,n}\delta_{k,j}} \right|^2 \\
&= \left| e^{-iE_{ph}(n+\frac{1}{2})t/\hbar}C_{n,j}\cos(\frac{\Omega_0t}{2})\right|^2 \\
&=|C_{n,j}|^2\cos^2(\frac{\Omega_0t}{2}).
\end{aligned}
\label{P1njt}
\end{equation}

\begin{figure}[h]
\begin{center}
\includegraphics[width=9cm]{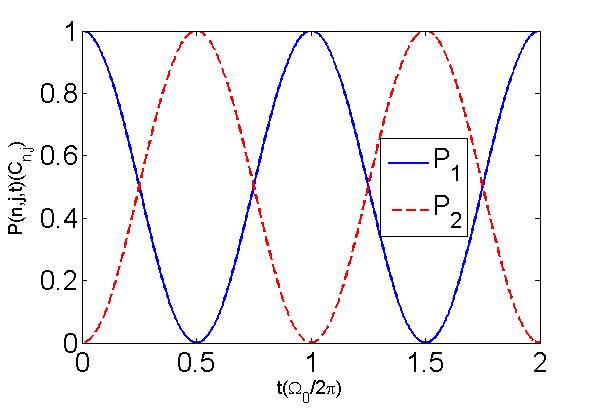}
\end{center}
\caption{Time evolution of a system between states $\spsi$ and $\spsf$. }
\label{Pnjt}
\end{figure}

Since $|C_{n,j}|^2$ is fixed for a given initial state of the system and can be normalized for the $(n,j)$ states, the probability of measuring $\vert \Phi_1^{(n,j)} \rangle$ state is proportional to $\cos^2(\frac{\Omega_0t}{2})$. Similarly, the probability of finding the system in an $\spsf$ state is
\begin{equation}
P_2(n,j,t)=|C_{n,j}|^2\sin^2(\frac{\Omega_0t}{2}).
\label{P2njt}
\end{equation}
From the two equations above, the Rabi angular frequency $\Omega_0$ holds for any $(n,j)$ states. The time evolution diagram for these two states is shown in Fig. \ref{Pnjt}.
These two dynamic evolution equations (Eqs. \ref{P1njt} and \ref{P2njt}) form the foundations of entanglement control technology.

For example, if at time $t=t_0$ we force the system into a known state, that is $C_{n,j}$ is known, then for subsequent times the system's states will evolve with a maximum amplitude described by $|C_{n,j}|^2$. If we try to excite the pure $m_s=j$ ground state to the excited state with a resonant laser pulse of length $\tau=\pi/\Omega_0$, then the system will evolve to the excited state at the end of the pulse. We call this kind of pulse a $\pi$-pulse. While if the pulse length $\tau=2\pi/\Omega_0$, it has no effect on the system. So, this kind of $2\pi$-pulse is a transparent pulse to the system. Generally, if $\tau$ is an arbitrary number, the pulse will apply an extra phase described by $\tau\Omega_0$ to the system. In field coupling application, the so called bang-bang coupling technique \cite{Morton2005} is also developed from this phenomenon.

\section{Electron structure of NV center in diamond}
To better control the quantum state of spin and accurately realize spin-photon entanglement, we need to consider the fine or even hyperfine level structures of the spinning electrons of the NV center (here, negatively charged) in the diamond crystal environment. There have been many approaches used to obtain the energy band structure, for example, Local Density Function method \cite{Gali2008},  many-body perturbation model \cite{Ma2010}, etc. In this section, I would like to calculate the matrix of the system's Hamiltonian with spin-spin and spin-orbit interaction, strain \cite{Fuchs2008}, \cite{Manson2006}, \cite{Rogers2009} terms. and use group theory's results to confirm the basis of electron states and transition selection rules with impact on the photon's polarization.

\begin{figure}[h]
\begin{center}
\includegraphics[width=7cm]{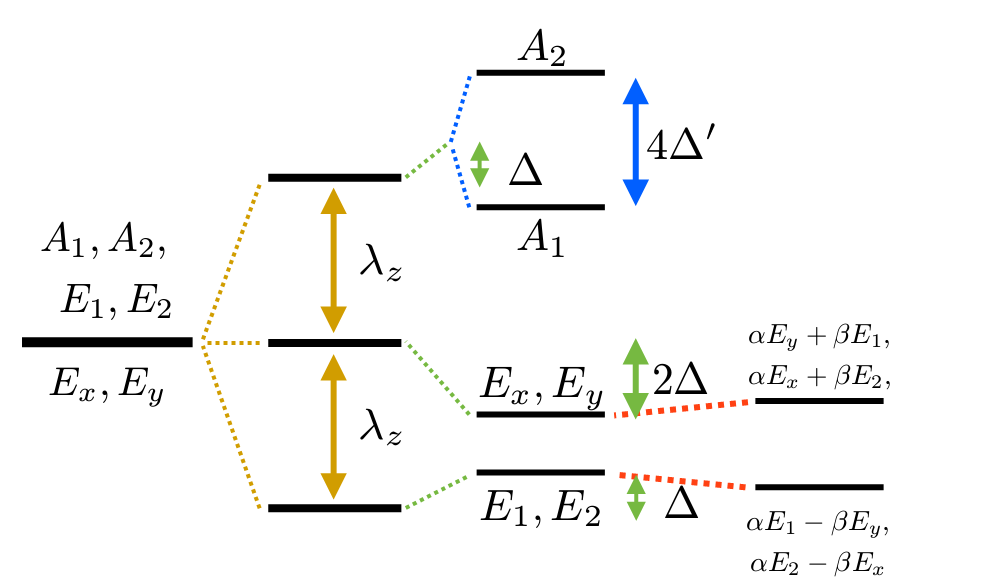}
\end{center}
\caption{Diagram of energy splitting of electrons in NV center, considering spin-orbit and spin-spin interactions.From J.R.~Maze, A.~Gali, E.~Togan, Y.~Chu, A.~Trifonov, E.~Kaxiras, and M.D.~Lukin,  Properties of nitrogen-vacancy centers in diamond: the group theoretic approach. \textit{New Journal of Physics}, 13 (2011), 025025 \href{http://dx.doi.org/10.1088/1367-2630/13/2/025025}{doi:10.1088/1367-2630/13/2/025025}\cite{Maze2011}. Reprinted with permission from IOP Publishing Ltd.}
\label{EnergySplitting1}
\end{figure}

In the absence of external strain and electric or magnetic fields,
properties of the six electronic excited states are determined by the NV center's $C_{3v}$ symmetry and spin-orbit and spin-spin interactions (shown in Fig.\ref{EnergySplitting1}). Optical transitions between the ground and excited states are spin preserving, but could change electronic orbital angular momentum depending on the photon polarization. Two of the excited states, labeled $\vert E_x\rangle$ and $\vert E_y\rangle$ according to their orbital symmetry,
correspond to the $m_s=0$ spin projection. Therefore they couple
only to the $\vert 0\rangle$ ground state and provide good cycling transitions, suitable for readout of the $\vert 0\rangle$ state population through fluorescence detection. The other four excited states are entangled states of spin and orbital angular momentum. Specifically, the $\vert A_2\rangle$ state has the form
\begin{equation}
\label{SA2}
\vert A_2 \rangle =\frac{1}{\sqrt{2}}\left( \vert E_-\rangle \vert +1 \rangle + \vert E_+\rangle \vert -1\rangle \right),
\end{equation}
where $\vert E_\pm\rangle$ are orbital states with angular momentum projection $\pm1$ along the NV axis, where $\vert \pm1 \rangle$ denotes the magnetic sublevel states with $m_s=\pm 1$. Similarly, we denote $m_s=0$ spin states with $\vert 0 \rangle$. At the same time, the ground states ($\vert 0\rangle, \vert \pm1\rangle$) are associated with the orbital state $\vert E_0\rangle$ with zero angular
momentum projection (for simplicity, the spatial part of the wavefunction is not explicitly written). Hence, owing to total angular momentum conservation, the $\vert A_2\rangle$ state decays with equal probability to the $\vert -1 \rangle$ ground
state through $\sigma_+$ polarized radiation and to $\vert +1\rangle$ through $\sigma_-$ polarized radiation (here, circular polarizations are represented by $\sigma_\pm=\hat{x} \pm \hat{y}$, while linear polarizations are represented by $\hat{x}$ and $\hat{y}$). In other words, if we can trigger a photon from the electron transition from $\vert A_2\rangle$ to the ground states $\vert \pm 1 \rangle$, we can entangle the frequency and spin state information into the polarized photon and easily verify the fidelity of entanglement by reading out the photon's polarization and frequency information and comparing this with our knowledge of the electron structure of the NV center. All of these properties make the $\vert A_2 \rangle$ singlet state (sometimes people use $\vert ^1 A_2 \rangle$ to differentiate this singlet state from the triplet state $\vert ^3 A_2\rangle$ which finally forms the singlet state under interaction with the atomic environment) a good candidate for spin-photon entangling in diamond NV center. And, it is vital to determine the level structure of related states before we carry out any entanglement scheme designs.

The inevitable presence of a small strain field, characterized by the strain splitting $\Delta_s$ of $\vert E_{x,y}\rangle$ reduces the NV center's symmetry and shifts the energies of the excited state ($\vert ^3 A_2\rangle$) levels according to their orbital wavefunctions. A group theory study of the NV center in diamond \cite{Maze2010}, tells us that a small strain will not change the polarization of the transited photon, such that we can use a small strain to modify the energy level structure. And group theory also gives the basis of electron states and the order of eigenvalues as well as the transition selection rule. The basis for the electron states of interest $\vert ^3 A_2\rangle$ gives \cite{Togan2010}
\begin{subequations}
\begin{align}
\vert A_1 \rangle &= \vert E_-\rangle\vert +1\rangle-\vert E_+\rangle\vert -1\rangle, \\
\vert A_2\rangle &= \vert E_-\rangle\vert +1\rangle+\vert E_+\rangle\vert -1\rangle, \\
\vert E_x \rangle &= \vert X \rangle \vert 0\rangle, \\
\vert E_y \rangle &= \vert Y \rangle \vert 0\rangle, \\
\vert E_1 \rangle &= \vert E_-\rangle\vert -1\rangle-\vert E_+\rangle\vert +1\rangle, \\
\vert A_2\rangle &= \vert E_-\rangle\vert -1\rangle+\vert E_+\rangle\vert +1\rangle,
\end{align}
\end{subequations}
where $\vert E_\pm \rangle=\vert ae_{\pm}-e_\pm a\rangle$, $\vert X(Y)\rangle=\vert ae_{x(y)}-e_{x(y)}a\rangle$, $e_\pm=\mp e_x-ie_y$ and $e_{x(y)}$, $a$ span into the full orbitals space of the electrons. Fig. \ref{EnergySplitting1} shows the energy splitting for electrons in NV center by considering spin-orbit and spin-spin interactions.

\begin{figure}[h]
\begin{center}
\includegraphics[width=13cm]{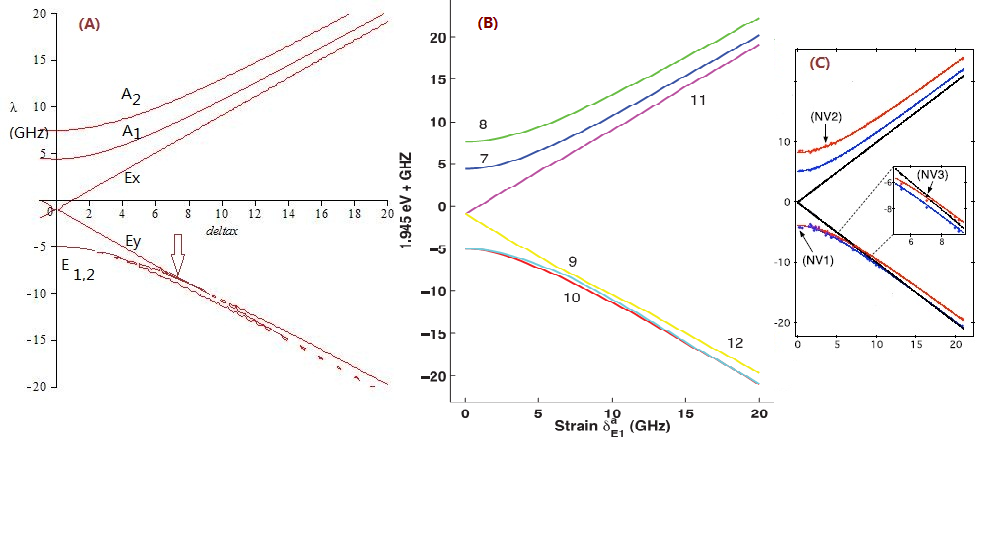}
\end{center}
\caption{Comparison of Electronic Structure as a function of stain. Equation~\ref{eigenstrain} leads to \textbf{(A)}, versus \cite{Maze2011} gives \textbf{(B)} (From \href{http://iopscience.iop.org/1367-2630/13/2/025025/}{J.R.~Maze, A.~Gali, E.~Togan, Y.~Chu, A.~Trifonov, E.~Kaxiras, and M.D.~Lukin,  Properties of nitrogen-vacancy centers in diamond: the group theoretic approach. \textit{New Journal of Physics}, 13 (2011), 025025}\cite{Maze2011}. Reprinted with permission from IOP Publishing Ltd.). The y-axis of Fig.(B) is moved down by 1.945 $eV$ to make the $E_x$ and $E_y$ originate from the origin. I have made $\delta_y=\delta_x=0.2$GHz in (A), which gives the similar lineshape as $\delta_{\perp}=7.3$GHz (only the slope changes slightly). The results of (A) and (B) agree very well, except for the crossing point under the arrow in (A). In (A), $E_2$ starts to surpass $E_y$ at the crossing point as the strain grows up. (A) much more close to the experimental result (yet disputable) shown in \textbf{(C)} (Reprinted figure
with permission from \href{http://prl.aps.org/abstract/PRL/v102/i19/e195506}{A.~Batalov, V.~Jacques, F.~Kaiser, P.~Siyushev, P.~Neumann, L.J.~Rogers, R.L.~McMurtrie, N.B.~Manson, F.~Jelezko, and J.~Wrachtrup, \textit{Phys. Rev. Lett.} 102, pp.195506 (2009)}\cite{Batalov2009}.  Copyright 2009 by the American Physical Society.), where the solid lines are calculated with the same set of parameters in (A), the discrete dots are measured from different NV centers label as $NV_1$, $NV_2$ and $NV_3$. The curves in (B) do not show the energy levels' cross between $E_2$ and $E_y$. }
\label{Strain}
\end{figure}

Including the strain effect, the total Hamiltonian reads
\begin{equation}
H = H_{ss}+H_{so}+H_{strain},
\end{equation}
where the spin-orbit interaction term $H_{so}$, spin-spin interaction term $H_{ss}$ and strain Hamiltonian are respectively given by \cite{Lenef1997}, \cite{Togan2010}
\begin{subequations}
\begin{align}
H_{so} &= \lambda_zS_zL_z+\lambda_\bot (LS)_\bot, \label{Hso}\\
H_{ss} &= \Delta \left( \vert A_1\rangle \langle A_1\vert+\vert A_2\rangle \langle A_2\vert
+\vert E_1\rangle \langle E_1 \vert +\vert E_2 \rangle \langle E_2 \vert \right) \\
 &\quad  -2\Delta \left( \vert E_x\rangle \langle E_x\vert+\vert E_y\rangle \langle E_y\vert\right)
+ \Delta'\left(\vert A_2\rangle \langle A_2 \vert -\vert A_1 \rangle \langle A_1 \vert \right), \\
H_{strain} &= \delta_x\left( \vert e_x\rangle \langle e_x\vert-\vert e_y\rangle \langle e_y\vert\right)
+\delta_y \left( \vert e_x\rangle \langle e_y\vert+\vert e_y\rangle \langle e_x\vert\right),
\end{align}
\end{subequations}
and $S$, $L$ are the magnetic and orbital angular operator, the subindex $\bot$ denotes the non-axial (or x-y plane) component. Following the procedure as before, we can rewrite the Hamiltonian into a matrix under the basis of $\left[A_2\quad A_1\quad E_x\quad E_y\quad E_2\quad E_1\right]$:
\begin{equation}
\mathbf{H}= \left( \begin {array}{cccccc} {\lambda_z}+\Delta+\Delta'&0&0&0&{\delta_y}&-{\delta_x}\\ \noalign{\medskip}0&{\lambda_z}+\Delta-\Delta'&0&0
&{\delta_x}&{\delta_y}\\
\noalign{\medskip}0&0&-2\Delta-{\delta_x}&
{\delta_y}&i{\lambda_{\bot}}&0\\
\noalign{\medskip}0&0&{\delta_y}&-2\Delta+{\delta_x}&0&i{\lambda_{\bot}}\\ \noalign{\medskip}{\delta_y}&{
\delta_x}&-i{\lambda_{\bot}}&0&-{\lambda_z}+\Delta&0\\ \noalign{\medskip}
-{\delta_x}&{\delta_y}&0&-i{\lambda_{\bot}}&0&-{\lambda_z}+\Delta
\end {array} \right),
\label{Hstrain}
\end{equation}
where $\Delta=1.42/3$GHz, $\Delta'=1.55$GHz, $\lambda_z=5.5$GHz, $\lambda_\bot=0.2$GHz \cite{Togan2010}. Notice that, upon the correspondence with Dr. J.R.~Maze, who is one of the coauthors of reference~\cite{Togan2010}, Eqs.~\ref{Hso} together with the matrix in Eq.~\ref{Hstrain} should be adjusted because the transverse spin-orbit does not mix the states with different spin projections on the excited state. And we should use $\lambda_\bot=7.3$GHz~\cite{Maze2011} as a result of \textit{ab initio} calculations instead of $0.2$GHz which is commonly used in the early work. To perform a comparison, we maintain $\lambda_\bot=0.2$GHz and continue to calculate the eigenenergy based on the equations above. If the eigenenergy is $\lambda$, then the eigenequation is given by
\begin{equation}
\begin{aligned}
&\quad {\lambda}^{6}- (3\,{{\delta_x}}^{2}+3\,{{\delta_y}}^{2}+64.32676667){\lambda}^{4}- 85.07314251\,{\lambda}^{3} \\
&+(3\,{{\delta_x}}^{4}+3\,{{\delta_y}}^{4}+6\,{{\delta_x}}^{2}
{{\delta_y}}^{2}+ 123.482\,({{\delta_x}}^{2}+ {{\delta_y}}^{2})+ 850.2580259){\lambda}^{2} \\
&+ (82.29690576\,({{\delta_x}}^{2} +\,{{\delta_y}}^{2})+ 1609.942968)\lambda -({{\delta_x}}^{6}+{{\delta_y}}^{6}) \\ &-3\,({{\delta_y}}^{2}{{\delta_x}}^{4}
 +\,{{\delta_y}}^{4}{{\delta_x}}^{2}) -59.15573333\,({{\delta_x}}^{4} +\,{{\delta_y}}^{4})- 118.3114667\,{{
\delta_x}}^{2}{{\delta_y}}^{2} \\
&- 0.1240\,{{\delta_x}}^{3}+ 0.3720\,{\delta_x}\,{{\delta_y}}^{2} - 787.4883241\,({{\delta_x}}^{2}+\,{{\delta_y}}^{2})+ 740.9385799=0. \label{eigenstrain}
 \end{aligned}
\end{equation}
I calculate the eigenenergies for states $A_1$, $A_2$, $E_x$, $E_y$, $E_2$ and $E_1$ using Maple. By forcing $\delta_y=\delta_x$ as a result of symmetry, I compared my result with Maze's result~\cite{Maze2011} and another independent result~\cite{Batalov2009}. They all agree with each other very well, except for a minor difference among the crossing points shown in Fig.~\ref{Strain}. As you can see, this minor difference does not affect the rest analysis in Togan's paper, since only one low strain value--far less than the cross point value--is used to realize the entanglement.

\begin{figure}[h]
\begin{center}
\includegraphics[width=13cm]{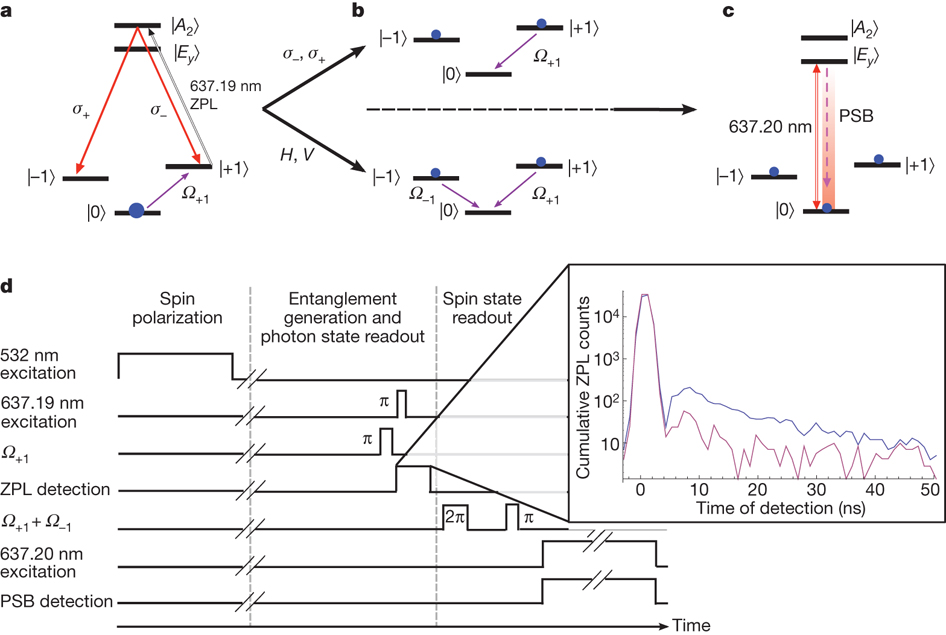}
\end{center}
\caption{\textbf{Electron level structure and experimental procedure for entanglement generation. a,} scheme of selected level structure used to generate entanglement: $\vert A_2\rangle$ state as the excited state and $\vert \pm 1\rangle$ states as the ground states. At the beginning, after spin polarization into $\vert 0\rangle$, population is transferred to $\vert +1\rangle$ by a microwave $\pi$-pulse ($\Omega_{+1}$). The NV is excited to $\vert A_2\rangle$ with a 637.19-nm $\pi$-pulse and the ZPL emission is collected. \textbf{b,} If a $\sigma_+$ or $\sigma_-$ photon is detected, the population in $\vert +1\rangle$ or $\vert -1\rangle$ is transferred to $\vert 0\rangle$. If an $\vert H\rangle$ or $\vert V\rangle$ photon is detected, a $\tau-2\pi-\tau$ echo sequence is applied with $\Omega_{+1}$ and $\Omega_{-1}$, followed by a $\pi$-pulse which transfers the population in $\vert M\rangle$ to $\vert 0\rangle$. \textbf{c,} The population in $\vert 0\rangle$ is measured using the 637.20-nm optical readout transition. \textbf{d,} Pulse sequence for the case where an $\vert H\rangle$ or $\vert V\rangle$ ZPL
photon is detected (time axis not to scale). If a $\sigma_{\pm}$ photon is detected instead, only a $\pi$-pulse on either $\Omega_{+1}$ or $\Omega_{-1}$ is used for spin readout. Inset, detection time of ZPL channel photons, showing reflection from diamond surface and subsequent NV emission (blue) and background counts (purple). Reprinted by permission from Macmillan Publishers Ltd: \href{http://www.nature.com/nature/index.html}{Nature} \textbf{466} (2010),
  no.~7307, 730--734\cite{Togan2010}, copyright 2010. }
\label{procedure}
\end{figure}

By introducing a low strain on the diamond, Togan and his teamworkers makes the energy gap between the $\vert A_2\rangle$ singlet state and the $\vert \pm 1\rangle$ states equal to the photon energy in the laser beam with a wavelength of about 637.19 nm, which can be easily obtained from a commonly used tunable YAG:Nd laser. The level scheme used to realize spin-light entanglement is a $\Lambda$-type system, shown in Fig. \ref{procedure}(a).

By the way, the so called $^{14}N$ and $^{13}C$ nucleus also contributes a hyperfine splitting to the electronic level structure in a diamond environment. To overcome this disadvantage,  we can use slightly detuned electronic microwaves ($\mu$ waves) to cover the effect of hyperfine coupling. This $\mu$ wave usually makes a detuning of $\Delta=120$ MHz, and makes an OSE energy shift of $\Omega_\mu/\Delta \approx 0.5$ MHz, considering the Rabi frequency for this hyperfine splitting is $8$ MHz. So, this effect is relatively small for the entanglement experiment. According to Eq.(\ref{dynamics}), by applying this $\mu$ wave, the ground states are rotated to a rotating frame described by
\begin{equation}
\vert \pm \! \tilde{1}\rangle_t=e^{i\phi_\pm}e^{-i(\omega_\pm \mp\Delta)t} \vert\pm \! 1\rangle.
\end{equation}
This rotation can easily transfer electrons to the $\vert 0 \rangle$ state from a superposition state
\begin{equation}
 \vert M\rangle=\frac{1}{\sqrt{2}}\left(e^{-i\omega_+t}\vert +1\rangle\pm e^{-i(\omega_-t-(\phi_+-\phi_-))}\vert -1\rangle \right),
\end{equation}
  corresponding to photon state $\vert H\rangle= \frac{1}{\sqrt{2}}(\vert \sigma_+\rangle +\vert \sigma_-\rangle)$ and $\vert V\rangle=\frac{1}{\sqrt{2}} (\vert \sigma_+\rangle -\vert \sigma_-\rangle)$, which are linear polarization states. In this way, it makes verification of non-diagram density matrix elements feasible (see Fig.\ref{statistics}(b)). The supplemental materiel of Togan's paper explained this technology.

\section{Decoding spin-photon entanglement experiments in NV center of diamond}
Now we are ready to design a spin-photon entanglement scheme for diamond NV centers.

As discussed before, if we can form a state described by
\begin{equation}
\vert \Psi \rangle = \frac{1}{\sqrt{2}}(\vert \sigma_-\rangle \vert +\!1\rangle+ \vert \sigma_+\rangle \vert -\!1\rangle),
\end{equation}
where $\vert \sigma \pm\rangle$ and $\vert \! \pm 1\!\rangle$ are the polarized photon and electronic spin states,
then we can say we have formed an entangled state. In August 2010, Togan and colleagues successfully realized this entangled state in NV centers in diamond nanocrystal and analyzed the entanglement in a basis of four Bell states. The electron levels and experimental procedures used to realize this entanglement are shown in Fig. \ref{procedure}. The authors used zero-phonon-line (ZPL) photons with four basis states: $\sigma_\pm$, $\vert H\rangle= \frac{1}{\sqrt{2}}(\vert \sigma_+\rangle +\vert \sigma_-\rangle)$ and $\vert V\rangle=\frac{1}{\sqrt{2}} (\vert \sigma_+\rangle -\vert \sigma_-\rangle)$. These photon states respectively entangled with four spin states: $\vert \pm \rangle$ and $\vert \pm \rangle = \frac{1}{\sqrt{2}}(\vert +1\rangle\pm \vert -1\rangle)$. These states span the four Bell states with maximum entanglement. It is worth mentioning that, to read out the photon and spin states, the authors used a carefully designed pulse echo sequence using the $\tau$-pulse, $\pi$-pulse and $2\pi$-pulse discussed in the on-resonance model. And as mentioned above, they also used a temporary state $\vert M\rangle$ in a rotating frame,  which can be explained using our off-resonant model. All of these technologies make detecting any expected states feasible.

\begin{figure}[h]
\begin{center}
\includegraphics[width=10cm]{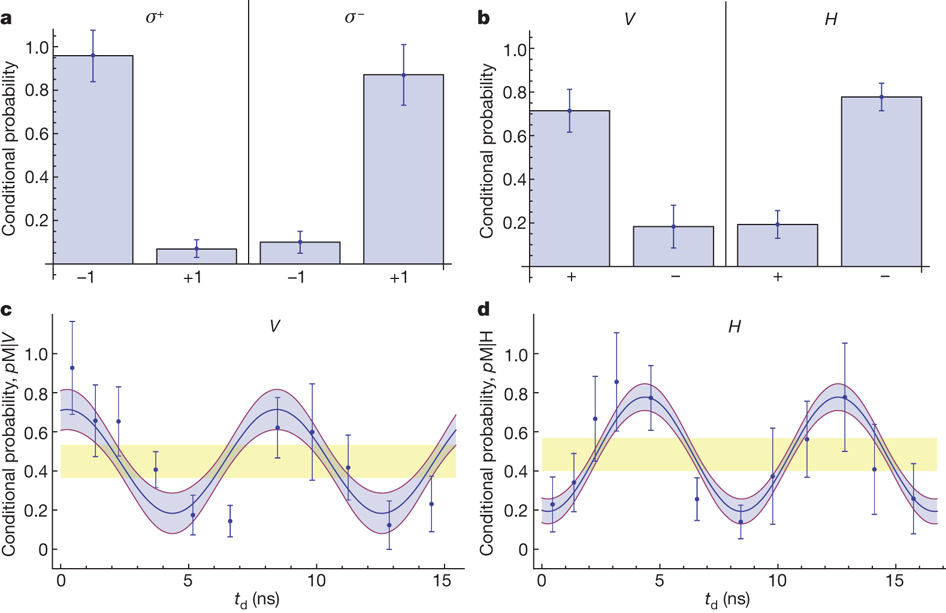}
\end{center}
\caption{\textbf{Measurement of spin-photon correlations in two bases. a,} Conditional probability of measuring $\vert \! \pm 1\!\rangle$ after the detection of a $\sigma_+$ or
 $\sigma_-$ photon. \textbf{b,} Conditional probability of measuring $\vert \! \pm 1\!\rangle$ after the detection
of an H or V photon, extracted from a fit to data shown in \textbf{c} and \textbf{d}.
\textbf{c, d,} Measured conditional probability of finding
the electronic spin in the state $\vert M \rangle$ after detection of
a \textbf{V}(\textbf{c}) or \textbf{H}(\textbf{d}) photon
at time $t_d$. Blue shaded region is the $68\%$ confidence interval for the fit (solid line) to the time binned
data. Errors bars on data points show $\pm1$ s.d. Combined with the data shown in \textbf{a}, oscillations with amplitude
outside of the yellow regions result in fidelities greater than $0.5$. The visibility
of the measured oscillations are $0.59\pm0.18$ (\textbf{c}) and $0.60\pm0.11$ (\textbf{d}). Reprinted by permission from Macmillan Publishers Ltd: \href{http://www.nature.com/nature/index.html}{Nature} \textbf{466} (2010),
  no.~7307, 730--734\cite{Togan2010}, copyright 2010.}
\label{statistics}
\end{figure}

Mainly because of the low detection rate, they spent several weeks to do one round of experiments to verify this entangled state, and carefully repeated the experiments over several months to collect enough data to publish their results. Finally, one beautiful statics diagram is shown in Fig.\ref{statistics}.

The probability of probing spin in $\vert -1 \rangle$ state after detecting a $\sigma^+$ photon is equivalent to $C(H,V)$ in Eq.\ref{Chv}, and so on. The probability of measuring the state $\vert M \rangle$ at time $t_d$ is
\begin{equation}
Pr_M(H,V,t_d)=\frac{1 \pm \cos(\alpha(t_d))}{2},
\end{equation}
where $\alpha(t_d)=(\omega_+-\omega_-)t_d+(\phi_+-\phi_-)$, with fixed $\phi_\pm$ as the initial condition for every experiment.
We can see the results give a very high fidelity for verifying the entanglement states.

\section{Conclusion and outlook}

In the approaches presented in this paper, we find the Jaynes-Cummings model can be successfully applied to analyze entanglement dynamics and technologies, including phase rotating, energy splitting (using strain or optical pump) and pulse echo technologies, in diamond NV centers, which is a promising medium for practical Quantum Informational and Quantum Computational applications. Our analysis of the electron structure also works well in this case. These approaches and technologies can be potentially used in similar materials and systems.

Through out the discussion, we can see that a well selected and controlled level structure is the key to realizing spin-photon entanglement. To make the spin information map onto photons, it is necessary to choose two spin opposite states and a common ground or excited state, which is the so-called $\Lambda$-type structure. In fact, some other level structures are also discussed in recent publications for realizing such an entanglement. For instance, a $W$-type structure is clearly discussed here \cite{Kosaka2009}. To fully understand this level design technology, a tomography analysis may be useful.

At the same time, to improve the entanglement performance, coupling the NV center mode to a high-Q cavity is also helpful (note from June, 2011: just reported by Patton and O'Brien that this enhancement was realized in a diamond microring~\cite{Patton2011}), especially to improve the entanglement distance and hence realize multi-qubit entanglement in solid-state materials. And since realizing entanglement networks is the final goal, it is necessary to study the coupling effects based on cavity-QED theories and discover practical quantum devices based on entangled units. To realize this goal, the symmetry of multiparticle entanglement \cite{Gour2010} and many-body dynamics theories \cite{Platzer2010} may also be helpful.

\section{Electronic Use Notice of Reprinted Materials}
As some of the figures presented in this essay are permitted and copyrighted by different publishers, the author of this paper would like to gratefully acknowledge the following statements, besides the credit lines adjacent to the figures.

Figs.~\ref{JClevel},~\ref{FE}~and~\ref{OSE} are licensed by AAAS. Fig.~\ref{Strain}(C) is permitted by American Physical Society. For these four figures: Readers may view, browse, and/or download material for temporary copying purposes only, provided these uses are for noncommercial personal purposes. Except as provided by law, this material may not be further reproduced, distributed, transmitted, modified, adapted, performed, displayed, published, or sold in whole or in part, without prior written permission from the publisher.

\section{Acknowledgements}
The author would like to thank Dr. Bob Buckley, Dr. Greg Fuchs, Dr. Lee Bassett, and Dr. David Awschalom for providing helpful clarifications and discussions on the work of Ref.~\cite{Buckley2010}, and thank Dr. J.R.~Maze for providing helpful discussion regarding the work of Ref.~\cite{Togan2010} and advice on using of materials in this essay. Xiaodong Qi would also like to thank Cole Van~Vlack and Emilia Illes for their help on the writing. Moreover, the author grateful acknowledge the authors and the publishers for the permission of reprinting some of the figures in this essay. Last but not least, the sincere gratitude from Xiaodong also goes to Marc Dignam for paying critical commitments on this piece of essay and leading Xiaodong into the exciting field of quantum optics and nonlinear optics through the course of study.

\bibliographystyle{amsplain}

\begin{thebibliography}{10}

\bibitem{barreiro2010experimental}
J.T. Barreiro, P.~Schindler, O.~G{\"u}hne, T.~Monz, M.~Chwalla, C.F. Roos,
  M.~Hennrich, and R.~Blatt, \emph{Experimental multiparticle entanglement
  dynamics induced by decoherence}, Nature Physics (2010).

\bibitem{Batalov2009}
A.~Batalov, V.~Jacques, F.~Kaiser, P.~Siyushev, P.~Neumann, L.~J. Rogers, R.~L.
  McMurtrie, N.~B. Manson, F.~Jelezko, and J.~Wrachtrup, \emph{Low temperature
  studies of the excited-state structure of negatively charged nitrogen-vacancy
  color centers in diamond}, Phys. Rev. Lett. \textbf{102} (2009), no.~19,
  195506.

\bibitem{Biercuk2009}
M.J. Biercuk, H.~Uys, A.P. VanDevender, N.~Shiga, W.M. Itano, and J.J.
  Bollinger, \emph{Optimized dynamical decoupling in a model quantum memory},
  Nature \textbf{458} (2009), no.~7241, 996--1000.

\bibitem{Buckley2010}
BB~Buckley, GD~Fuchs, LC~Bassett, and DD~Awschalom, \emph{Spin-light coherence
  for single-spin measurement and control in diamond}, Science \textbf{330}
  (2010), no.~6008, 1212.

\bibitem{Bukach2010}
AA~Bukach and S.Y. Kilin, \emph{Creation of entangled state between two spaced
  nv centers in diamond}, Optics and Spectroscopy \textbf{108} (2010), no.~2,
  254--266.

\bibitem{Chen2010b}
Qiong Chen, Zhenyu Xu, and Mang Feng, \emph{Entanglement generation of
  nitrogen-vacancy centers via coupling to nanometer-sized resonators and a
  superconducting interference device}, Phys. Rev. A \textbf{82} (2010), no.~1,
  014302.

\bibitem{Cohen2009}
Offir Cohen, Jeff~S. Lundeen, Brian~J. Smith, Graciana Puentes, Peter~J.
  Mosley, and Ian~A. Walmsley, \emph{Tailored photon-pair generation in optical
  fibers}, Phys. Rev. Lett. \textbf{102} (2009), no.~12, 123603.

\bibitem{Dousse2010}
A.~Dousse, J.~Suffczynski, O.~Krebs, A.~Beveratos, A.~Lemaitre, I.~Sagnes,
  J.~Bloch, P.~Voisin, and P.~Senellart, \emph{A quantum dot based bright
  source of entangled photon pairs operating at 53 k}, Applied Physics Letters
  \textbf{97} (2010), no.~8, 081104.

\bibitem{Du2009}
J.~Du, X.~Rong, N.~Zhao, Y.~Wang, J.~Yang, and RB~Liu, \emph{Preserving
  electron spin coherence in solids by optimal dynamical decoupling}, Nature
  \textbf{461} (2009), no.~7268, 1265--1268.

\bibitem{Eisaman2008}
M.~D. Eisaman, E.~A. Goldschmidt, J.~Chen, J.~Fan, and A.~Migdall,
  \emph{Experimental test of nonlocal realism using a fiber-based source of
  polarization-entangled photon pairs}, Phys. Rev. A \textbf{77} (2008), no.~3,
  032339.

\bibitem{Eisenberg2004}
H.~S. Eisenberg, G.~Khoury, G.~A. Durkin, C.~Simon, and D.~Bouwmeester,
  \emph{Quantum entanglement of a large number of photons}, Phys. Rev. Lett.
  \textbf{93} (2004), no.~19, 193901.

\bibitem{Eisert2005}
J.~Eisert and D.~Gross, \emph{Multi-particle entanglement}, Arxiv preprint
  quant-ph/0505149 (2005).

\bibitem{Eisert2003}
J.~Eisert and MB~Plenio, \emph{Introduction to the basics of entanglement
  theory in continuous-variable systems}, Arxiv preprint quant-ph/0312071
  (2003).

\bibitem{Eisert2005a}
J.~Eisert and MM~Wolf, \emph{Gaussian quantum channels}, Arxiv preprint
  quant-ph/0505151 (2005).

\bibitem{Fan2007}
J.~Fan, M.~D. Eisaman, and A.~Migdall, \emph{Bright phase-stable broadband
  fiber-based source of polarization-entangled photon pairs}, Phys. Rev. A
  \textbf{76} (2007), no.~4, 043836.

\bibitem{Faraon2011}
A.~Faraon, P.E. Barclay, C.~Santori, K.M.C. Fu, and R.G. Beausoleil,
  \emph{Resonant enhancement of the zero-phonon emission from a colour centre
  in a diamond cavity}, Nature Photonics (2011).

\bibitem{Fuchs2008}
G.~D. Fuchs, V.~V. Dobrovitski, R.~Hanson, A.~Batra, C.~D. Weis, T.~Schenkel,
  and D.~D. Awschalom, \emph{Excited-state spectroscopy using single spin
  manipulation in diamond}, Phys. Rev. Lett. \textbf{101} (2008), no.~11,
  117601.

\bibitem{Gali2009}
Adam Gali, \emph{Theory of the neutral nitrogen-vacancy center in diamond and
  its application to the realization of a qubit}, Phys. Rev. B \textbf{79}
  (2009), no.~23, 235210.

\bibitem{Gali2008}
Adam Gali, Maria Fyta, and Efthimios Kaxiras, \emph{Ab initio supercell
  calculations on nitrogen-vacancy center in diamond: Electronic structure and
  hyperfine tensors}, Phys. Rev. B \textbf{77} (2008), no.~15, 155206.

\bibitem{Gerry2005}
C.C. Gerry and P.L. Knight, \emph{Introductory quantum optics}, Cambridge Univ
  Pr, 2005.

\bibitem{Gonzalez2010}
Gabriel González and Michael~N Leuenberger, \emph{The dynamics of the
  optically driven Λ transition of the 15 n–v − center in diamond},
  Nanotechnology \textbf{21} (2010), no.~27, 274020.

\bibitem{Goss1997}
J.~P. Goss, R.~Jones, P.~R. Briddon, G.~Davies, A.~T. Collins, A.~Mainwood,
  J.~A. van Wyk, J.~M. Baker, M.~E. Newton, A.~M. Stoneham, and S.~C. Lawson,
  \emph{Comment on ``electronic structure of the n-$v$ center in diamond:
  Theory''}, Phys. Rev. B \textbf{56} (1997), no.~24, 16031--16032.

\bibitem{Gour2010}
Gilad Gour, \emph{Evolution and symmetry of multipartite entanglement}, Phys.
  Rev. Lett. \textbf{105} (2010), no.~19, 190504.

\bibitem{Gregg2007}
J.F. Gregg, \emph{Spintronics: A growing science}, Nature Materials \textbf{6}
  (2007), no.~11, 798--799.

\bibitem{Hemmer2005}
P.~Hemmer and J.~Wrachtrup, \emph{Where is my quantum computer?}, Phys. Rev. A
  \textbf{72} (2005), 052330.

\bibitem{Herrmann2010}
LG~Herrmann, F.~Portier, P.~Roche, AL~Yeyati, T.~Kontos, and C.~Strunk,
  \emph{Carbon nanotubes as cooper-pair beam splitters.}, Physical review
  letters \textbf{104} (2010), no.~2, 026801.

\bibitem{Hossain2008}
Faruque~M. Hossain, Marcus~W. Doherty, Hugh~F. Wilson, and Lloyd C.~L.
  Hollenberg, \emph{Ab initio electronic and optical properties of the
  $n-v^{-}$ center in diamond}, Phys. Rev. Lett. \textbf{101} (2008), no.~22,
  226403.

\bibitem{Kosaka2009}
H.~Kosaka, T.~Inagaki, Y.~Rikitake, H.~Imamura, Y.~Mitsumori, and K.~Edamatsu,
  \emph{Spin state tomography of optically injected electrons in a
  semiconductor}, Nature \textbf{457} (2009), no.~7230, 702--705.

\bibitem{Ladd2010}
TD~Ladd, F.~Jelezko, R.~Laflamme, Y.~Nakamura, C.~Monroe, and JL~O’Brien,
  \emph{Quantum computers}, Nature \textbf{464} (2010), no.~7285, 45--53.

\bibitem{Lambrecht2010}
W.R.L. Lambrecht, \emph{Which electronic structure method for the study of
  defects: A commentary}, physica status solidi (b) (2010).

\bibitem{Larsson2008}
J.~A. Larsson and P.~Delaney, \emph{Electronic structure of the
  nitrogen-vacancy center in diamond from first-principles theory}, Phys. Rev.
  B \textbf{77} (2008), no.~16, 165201.

\bibitem{Lee2009}
J.~D. Lee, H.~Gomi, and Muneaki Hase, \emph{Coherent optical control of the
  ultrafast dephasing and mobility in a polar semiconductor}, Journal of
  Applied Physics \textbf{106} (2009), no.~8, 083501.

\bibitem{Lee2008}
J.~D. Lee and Muneaki Hase, \emph{Coherent optical control of the ultrafast
  dephasing of phonon-plasmon coupling in a polar semiconductor using a pulse
  train of below-band-gap excitation}, Phys. Rev. Lett. \textbf{101} (2008),
  no.~23, 235501.

\bibitem{Lenef1997}
A.~Lenef and S.~C. Rand, \emph{Reply to ``comment on `electronic structure of
  the n-$v$ center in diamond: Theory' ''}, Phys. Rev. B \textbf{56} (1997),
  no.~24, 16033--16034.

\bibitem{Liang1999}
Y.~Liang, JW~Lou, JK~Andersen, JC~Stocker, O.~Boyraz, MN~Islam, and DA~Nolan,
  \emph{Polarization-insensitive nonlinear optical loop mirror demultiplexer
  with twisted fiber}, Optics letters \textbf{24} (1999), no.~11, 726--728.

\bibitem{Liu2009b}
Jiang-Tao Liu, Fu-Hai Su, and Hai Wang, \emph{Model of the optical stark effect
  in semiconductor quantum wells: Evidence for asymmetric dressed exciton
  bands}, Phys. Rev. B \textbf{80} (2009), no.~11, 113302.

\bibitem{Ma2010}
Yuchen Ma, Michael Rohlfing, and Adam Gali, \emph{Excited states of the
  negatively charged nitrogen-vacancy color center in diamond}, Phys. Rev. B
  \textbf{81} (2010), no.~4, 041204.

\bibitem{Majer2007}
J.~Majer, JM~Chow, JM~Gambetta, J.~Koch, BR~Johnson, JA~Schreier, L.~Frunzio,
  DI~Schuster, AA~Houck, A.~Wallraff, et~al., \emph{Coupling superconducting
  qubits via a cavity bus}, Nature \textbf{449} (2007), no.~7161, 443--447.

\bibitem{Manson2006}
N.~B. Manson, J.~P. Harrison, and M.~J. Sellars, \emph{Nitrogen-vacancy center
  in diamond: Model of the electronic structure and associated dynamics}, Phys.
  Rev. B \textbf{74} (2006), no.~10, 104303.

\bibitem{mason2010carbon}
N.~Mason, \emph{Carbon nanotubes help pairs survive a breakup},  (2010).

\bibitem{Maze2011}
J.R.~Maze, A.~Gali, E.~Togan, Y.~Chu, A.~Trifonov, E.~Kaxiras, and M.D.~Lukin,  \emph{Properties of nitrogen-vacancy centers in diamond: the group theoretic approach}. New Journal of Physics \textbf{13} (2011), 025025,

\bibitem{Milburn2010}
G.J. Milburn, \emph{Quantum measurement and control of single spins in
  diamond}, Science \textbf{330} (2010), no.~6008, 1188.

\bibitem{Morton2005}
J.J.L. Morton, A.M. Tyryshkin, A.~Ardavan, S.C. Benjamin, K.~Porfyrakis,
  SA~Lyon, and G.A.D. Briggs, \emph{Bang--bang control of fullerene qubits
  using ultrafast phase gates}, Nature Physics \textbf{2} (2005), no.~1,
  40--43.

\bibitem{Muller2009}
Andreas Muller, Wei Fang, John Lawall, and Glenn~S. Solomon, \emph{Creating
  polarization-entangled photon pairs from a semiconductor quantum dot using
  the optical stark effect}, Phys. Rev. Lett. \textbf{103} (2009), no.~21,
  217402.

\bibitem{Olsson2000}
B.E. Olsson, P.~Ohlen, L.~Rau, and D.J. Blumenthal, \emph{A simple and robust
  40-gb/s wavelength converter using fiber cross-phase modulation and optical
  filtering}, Photonics Technology Letters, IEEE \textbf{12} (2000), no.~7,
  846--848.

\bibitem{Patton2011}
B.R. Patton and J.L. O'Brien, \emph{Integrated quantum photonics: Photons in a
  diamond microring}, Nature Photonics \textbf{5} (2011), no.~5, 256--258.

\bibitem{Platzer2010}
Felix Platzer, Florian Mintert, and Andreas Buchleitner, \emph{Optimal
  dynamical control of many-body entanglement}, Phys. Rev. Lett. \textbf{105}
  (2010), no.~2, 020501.

\bibitem{Plenio2005}
M.B. Plenio and S.~Virmani, \emph{An introduction to entanglement measures},
  Arxiv preprint quant-ph/0504163 (2005).

\bibitem{Raadmark2009a}
M.~R{\aa}dmark, M.~{\.Z}ukowski, and M.~Bourennane, \emph{Experimental high
  fidelity six-photon entangled state for telecloning protocols}, New Journal
  of Physics \textbf{11} (2009), 103016.

\bibitem{Ramsay2010}
A~J Ramsay, \emph{A review of the coherent optical control of the exciton and
  spin states of semiconductor quantum dots}, Semiconductor Science and
  Technology \textbf{25} (2010), no.~10, 103001.

\bibitem{Maze2010}
J.R. Maze, \emph{Quantum manipulation of nitrogen-vacancy centers in diamond:
  from basic properties to applications}, Ph.D. thesis, Harvard University
  Cambridge, Massachusetts, 2010.

\bibitem{Robledo2010}
L.~Robledo, H.~Bernien, T.~van~der Sar, and R.~Hanson, \emph{Spin dynamics in
  the optical cycle of single nitrogen-vacancy centres in diamond}, Arxiv
  preprint arXiv:1010.1192 (2010).

\bibitem{Rogers2009}
L~J Rogers, R~L McMurtrie, M~J Sellars, and N~B Manson, \emph{Time-averaging
  within the excited state of the nitrogen-vacancy centre in diamond}, New
  Journal of Physics \textbf{11} (2009), no.~6, 063007.

\bibitem{Santori2010}
C.~Santori, PE~Barclay, KM~Fu, RG~Beausoleil, S.~Spillane, and M.~Fisch,
  \emph{Nanophotonics for quantum optics using nitrogen-vacancy centers in
  diamond}, Nanotechnology \textbf{21} (2010), 274008.

\bibitem{Sarchi2008}
D.~Sarchi and V.~Savona, \emph{Spectrum and thermal fluctuations of a
  microcavity polariton bose-einstein condensate}, Phys. Rev. B \textbf{77}
  (2008), 045304.

\bibitem{Sarma2001}
S.~Das Sarma, Jaroslav Fabian, Xuedong Hu, and Igor~Z[combining breve]utic,
  \emph{Spin electronics and spin computation}, Solid State Communications
  \textbf{119} (2001), no.~4-5, 207 -- 215.

\bibitem{Sotier2009a}
F.~Sotier, T.~Thomay, T.~Hanke, J.~Korger, S.~Mahapatra, A.~Frey, K.~Brunner,
  R.~Bratschitsch, and A.~Leitenstorfer, \emph{Femtosecond few-fermion dynamics
  and deterministic single-photon gain in a quantum dot}, Nature Physics
  \textbf{5} (2009), no.~5, 352--356.

\bibitem{Stoneham2009}
A~Marshall Stoneham, A~H Harker, and Gavin~W Morley, \emph{Could one make a
  diamond-based quantum computer?}, Journal of Physics: Condensed Matter
  \textbf{21} (2009), no.~36, 364222.

\bibitem{Tamarat2008}
Ph~Tamarat, N~B Manson, J~P Harrison, R~L McMurtrie, A~Nizovtsev, C~Santori,
  R~G Beausoleil, P~Neumann, T~Gaebel, F~Jelezko, P~Hemmer, and J~Wrachtrup,
  \emph{Spin-flip and spin-conserving optical transitions of the
  nitrogen-vacancy centre in diamond}, New Journal of Physics \textbf{10}
  (2008), no.~4, 045004.

\bibitem{Togan2010}
E.~Togan, Y.~Chu, AS~Trifonov, L.~Jiang, J.~Maze, L.~Childress, MVG Dutt, A.S.
  S{\o}rensen, PR~Hemmer, AS~Zibrov, et~al., \emph{Quantum entanglement between
  an optical photon and a solid-state spin qubit}, Nature \textbf{466} (2010),
  no.~7307, 730--734.

\bibitem{Ulrich1979}
R.~Ulrich and A.~Simon, \emph{Polarization optics of twisted single-mode
  fibers}, Appl. Opt. \textbf{18} (1979), no.~13, 2241--2251.

\bibitem{Unold2004}
Thomas Unold, Kerstin Mueller, Christoph Lienau, Thomas Elsaesser, and
  Andreas~D. Wieck, \emph{Optical stark effect in a quantum dot: Ultrafast
  control of single exciton polarizations}, Phys. Rev. Lett. \textbf{92}
  (2004), no.~15, 157401.

\bibitem{Loock2000}
P.~van Loock and Samuel~L. Braunstein, \emph{Multipartite entanglement for
  continuous variables: A quantum teleportation network}, Phys. Rev. Lett.
  \textbf{84} (2000), no.~15, 3482--3485.

\bibitem{Vidal2002}
G.~Vidal and R.~F. Werner, \emph{Computable measure of entanglement}, Phys.
  Rev. A \textbf{65} (2002), no.~3, 032314.

\bibitem{Vu2006}
Q.~T. Vu, H.~Haug, and S.~W. Koch, \emph{Relaxation and dephasing quantum
  kinetics for a quantum dot in an optically excited quantum well}, Phys. Rev.
  B \textbf{73} (2006), no.~20, 205317.

\bibitem{Weber2010}
JR~Weber, WF~Koehl, JB~Varley, A.~Janotti, BB~Buckley, CG~Van~de Walle, and
  DD~Awschalom, \emph{Quantum computing with defects}, Proceedings of the
  National Academy of Sciences \textbf{107} (2010), no.~19, 8513.

\bibitem{Wolters2010}
Janik Wolters, Andreas~W. Schell, Gunter Kewes, Nils Nusse, Max Schoengen,
  Henning Doscher, Thomas Hannappel, Bernd Lochel, Michael Barth, and Oliver
  Benson, \emph{Enhancement of the zero phonon line emission from a single
  nitrogen vacancy center in a nanodiamond via coupling to a photonic crystal
  cavity}, Applied Physics Letters \textbf{97} (2010), no.~14, 141108.

\bibitem{Yang2010a}
Wanli Yang, Zhenyu Xu, Mang Feng, and Jiangfeng Du, \emph{Entanglement of
  separate nitrogen-vacancy centers coupled to a whispering-gallery mode
  cavity}, New Journal of Physics \textbf{12} (2010), no.~11, 113039.

\bibitem{Yang2010b}
WL~Yang, ZQ~Yin, ZY~Xu, M.~Feng, and JF~Du, \emph{One-step implementation of
  multiqubit conditional phase gating with nitrogen-vacancy centers coupled to
  a high-q silica microsphere cavity}, Applied Physics Letters \textbf{96}
  (2010), no.~24, 241113.

\bibitem{Yoo2010}
J.W. Yoo, C.Y. Chen, HW~Jang, CW~Bark, VN~Prigodin, CB~Eom, and AJ~Epstein,
  \emph{Spin injection/detection using an organic-based magnetic
  semiconductor}, Nature materials \textbf{9} (2010), no.~8, 638--642.

\end{thebibliography}
\providecommand{\bysame}{\leavevmode\hbox to3em{\hrulefill}\thinspace}
\providecommand{\MR}{\relax\ifhmode\unskip\space\fi MR }
\providecommand{\MRhref}[2]{%
  \href{http://www.ams.org/mathscinet-getitem?mr=#1}{#2}
}
\providecommand{\href}[2]{#2}


\end{document}